\newlength{\newhcol}
\documentclass[10pt,twocolumn,twoside]{IEEEtran}
\newcommand{\keyitem}[1]{\vspace{9pt}\noindent\textbf{#1:}}

\newcommand\MYhyperrefoptions{bookmarks=false,bookmarksnumbered=false,
pdfpagemode={none},plainpages=false,pdfpagelabels=true,
colorlinks=true,linkcolor={black},citecolor={black},pagecolor={black},
urlcolor={black},
breaklinks={true},
pdftitle={},
pdfsubject={},
pdfauthor={},
pdfkeywords={}}

\usepackage[\MYhyperrefoptions,pdftex]{hyperref}

\usepackage{algpseudocode}
\usepackage{algorithm}
\usepackage{fmtcount}
\usepackage{mdwlist}
\usepackage{textcomp}
\usepackage{slashbox}

\usepackage[caption=false,font=footnotesize]{subfig}
\usepackage{amssymb}
\usepackage{amsthm}
\usepackage{latexsym}
\usepackage{verbatim}
\usepackage[pdftex]{graphicx}
\usepackage{fixltx2e}
\usepackage{array}
\usepackage{booktabs}
\usepackage[square, numbers,sort&compress]{natbib}
\usepackage[cmex10]{amsmath}
\usepackage{setspace,algorithm,enumerate}
\usepackage{array}

\usepackage{amsmath}
\usepackage{float}
\usepackage{paralist}

\newtheorem{theorem}{Theorem}

{\begin{list}               
    {$\bullet$ \hfill}{
        \setlength{\leftmargin}{\parindent}
        \setlength{\parsep}{0.04\baselineskip}
        \setlength{\itemsep}{0.5\parsep}
        \setlength{\labelwidth}{\leftmargin}
        \setlength{\labelsep}{0em}}
    }
{\end{list}}

\providecommand{\cref}[1]{Chapter~\ref{chap:#1}}

\providecommand{\R}{\ensuremath{\mathbb{R}}}

\renewcommand{\vec}[1]{\ensuremath{\boldsymbol{#1}}}
\providecommand{\mat}[1]{\ensuremath{\boldsymbol{#1}}}

\providecommand{\calS}{\mathcal{S}}

\providecommand{\mA}{\mat{A}} 
\providecommand{\mC}{\mat{C}} 
 
\providecommand{\mI}{\mat{I}}

\providecommand{\mPhi}{{\mat{\Phi}}}
\providecommand{\mPsi}{{\mat{\Psi}}}
\providecommand{\mpsi}{{\mat{\psi}}}
\providecommand{\mtau}{{\mat{\tau}}}

\providecommand{\valpha}{{\mat{\alpha}}}

\providecommand{\vomega}{{\mat{\omega}}}

 \providecommand{\vp}{\vec{p}}
 
\providecommand{\vs}{\vec{s}}

\providecommand{\vx}{\vec{x}} \providecommand{\vy}{\vec{y}}

\newcommand\meteoswiss{{MeteoSwiss}}
\newcommand\meteoswissref{~\cite{meteoswiss}}
\newcommand\dataA{\emph{Payerne}}
\newcommand\dataB{\emph{Valais}}
\newcommand\sensorscope{\cite{Ingelrest2010}}

\begin{document}

\title{DASS: Distributed Adaptive Sparse Sensing}
\author{
\IEEEauthorblockN{Zichong Chen, Juri Ranieri, Runwei Zhang, and Martin
  Vetterli}
\thanks{The results of this research are reproducible: The datasets and Matlab codes used to generate figures can be found in our reproducible repository at \url{http://rr.epfl.ch/}. This research is supported by Swiss National Centre of Competence in Research and ERC Advanced Investigators Grant of European Union.}
\thanks{Z. Chen, J. Ranieri, R. Zhang and M. Vetterli are with the LCAV, I\&C, \'{E}cole Polytechnique F\'{e}d\'{e}rale de Lausanne (EPFL), Lausanne, Switzerland (e-mail: chenzc04@gmail.com, juri.ranieri@epfl.ch, runwei.zhang@epfl.ch, martin.vetterli@epfl.ch).}}

\maketitle

\begin{abstract}
  Wireless sensor networks are often designed to perform two
  tasks: sensing a physical field and transmitting the data to end-users.
  A crucial aspect of the design of a WSN is the minimization of the
  overall energy consumption. Previous researchers aim at optimizing the
  energy spent for the communication, while mostly ignoring the energy cost
  due to sensing.

  Recently, it has been shown that considering the sensing energy cost
  can be beneficial for further improving the overall energy
  efficiency. More precisely, sparse sensing techniques were proposed
  to reduce the amount of collected samples and
  recover the missing data by using data statistics. While the
  majority of these techniques use fixed or random sampling
  patterns, we propose to adaptively learn the signal
  model from the measurements and use the model to schedule when and
  where to sample the physical field.

  The proposed method requires minimal on-board computation, no
  inter-node communications and still achieves appealing reconstruction
  performance.  With experiments on real-world datasets, we
  demonstrate significant improvements over both traditional sensing
  schemes and the state-of-the-art sparse sensing schemes, particularly when the measured data is characterized by a strong intra-sensor (temporal) or inter-sensors (spatial) correlation.
\end{abstract}

\section{Introduction}
\label{sec1}

In a wireless sensor network (WSN), sensor nodes are deployed to take
periodical measurements of a certain physical field at different
locations. Consider a continuous-time spatio-temporal field $x(\vp,t)$
that we would like to monitor with the WSN and a vector $\vx\in\R^{N}$
containing a discretization of such field with a sufficiently high
resolution for our purposes. The target of the WSN is to
recover $\vx$ with the maximum precision.

\begin{figure*}[!t]
\centering
\subfloat[Traditional Sensing]{\label{fig:eye}\includegraphics[height=0.135\textheight]{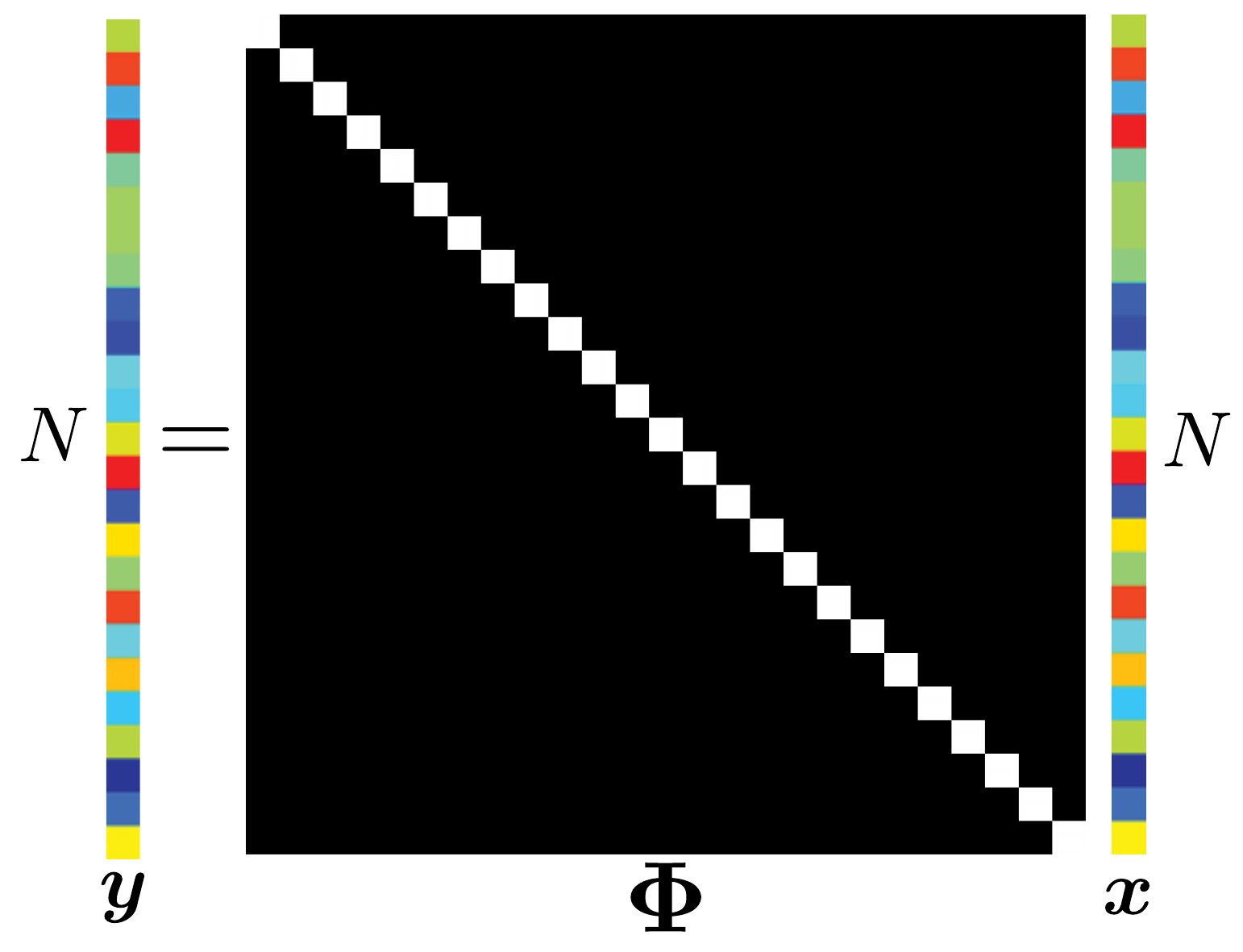}}
\subfloat[CS - Dense
Matrix]{\label{fig:dense}\includegraphics[height=0.135\textheight]{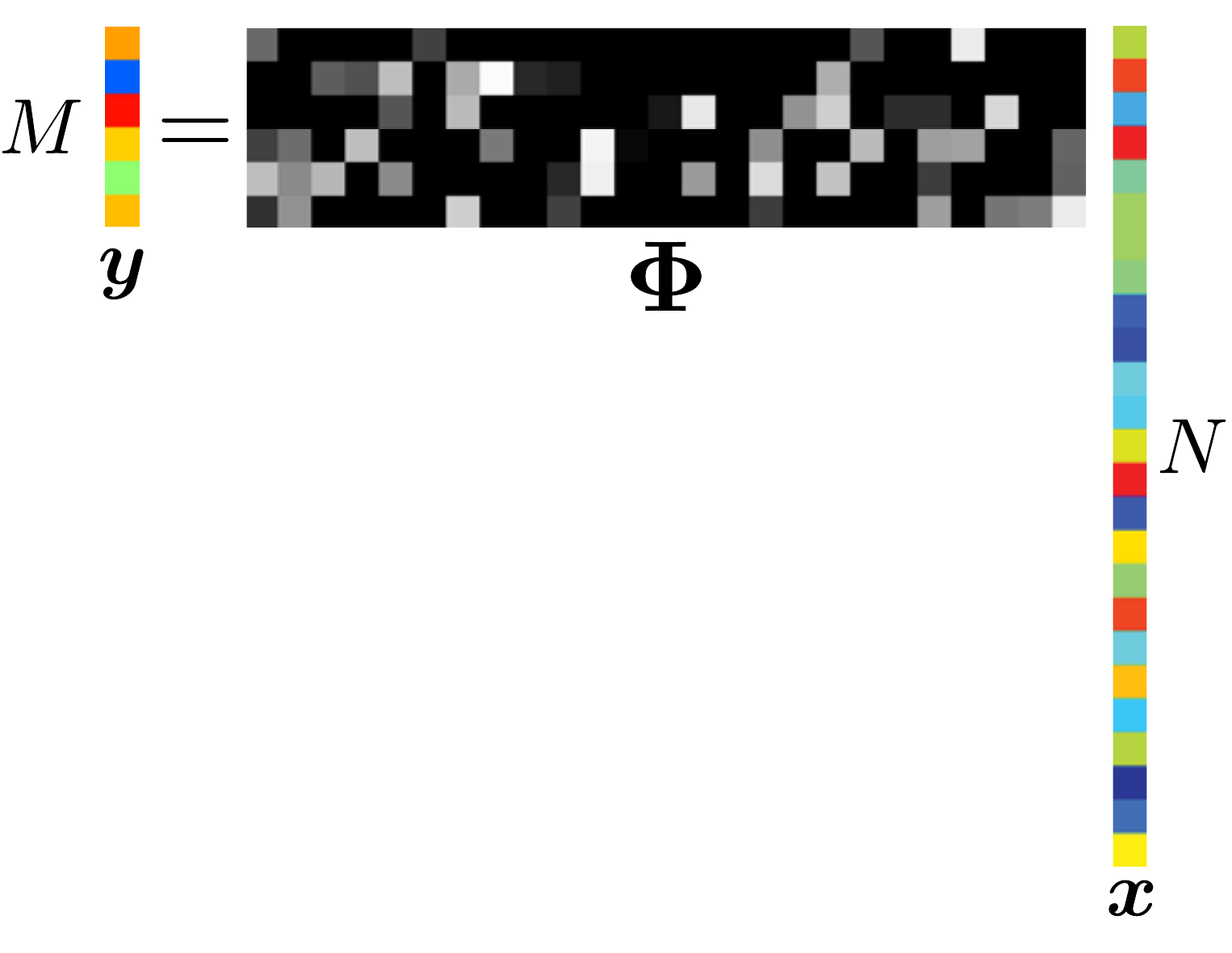}}
\subfloat[CS - Sparse
Matrix]{\label{fig:sparse}\includegraphics[height=0.135\textheight]{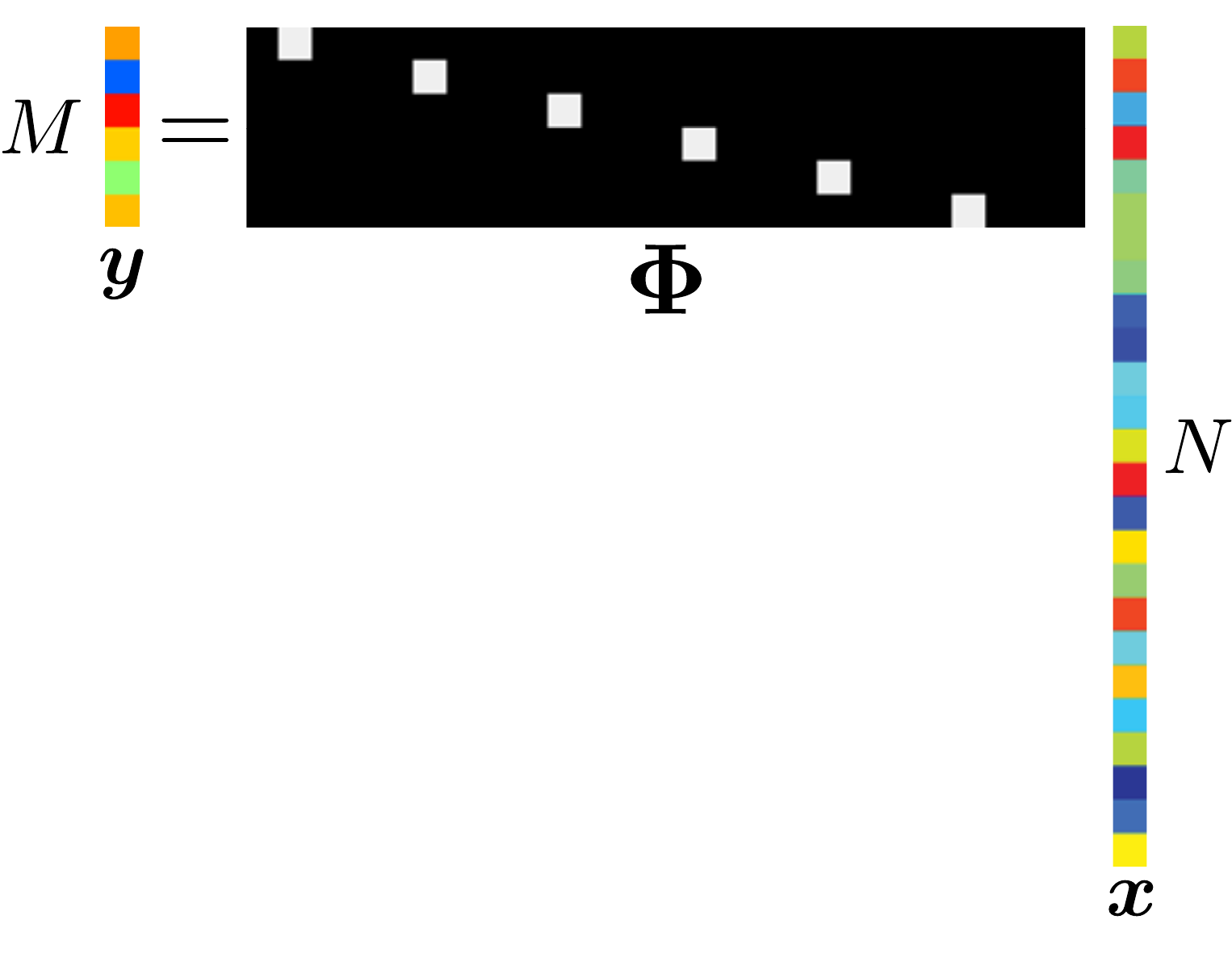}}
\subfloat[Sparsity Dictionary]{\label{fig:basis}\includegraphics[height=0.135\textheight]{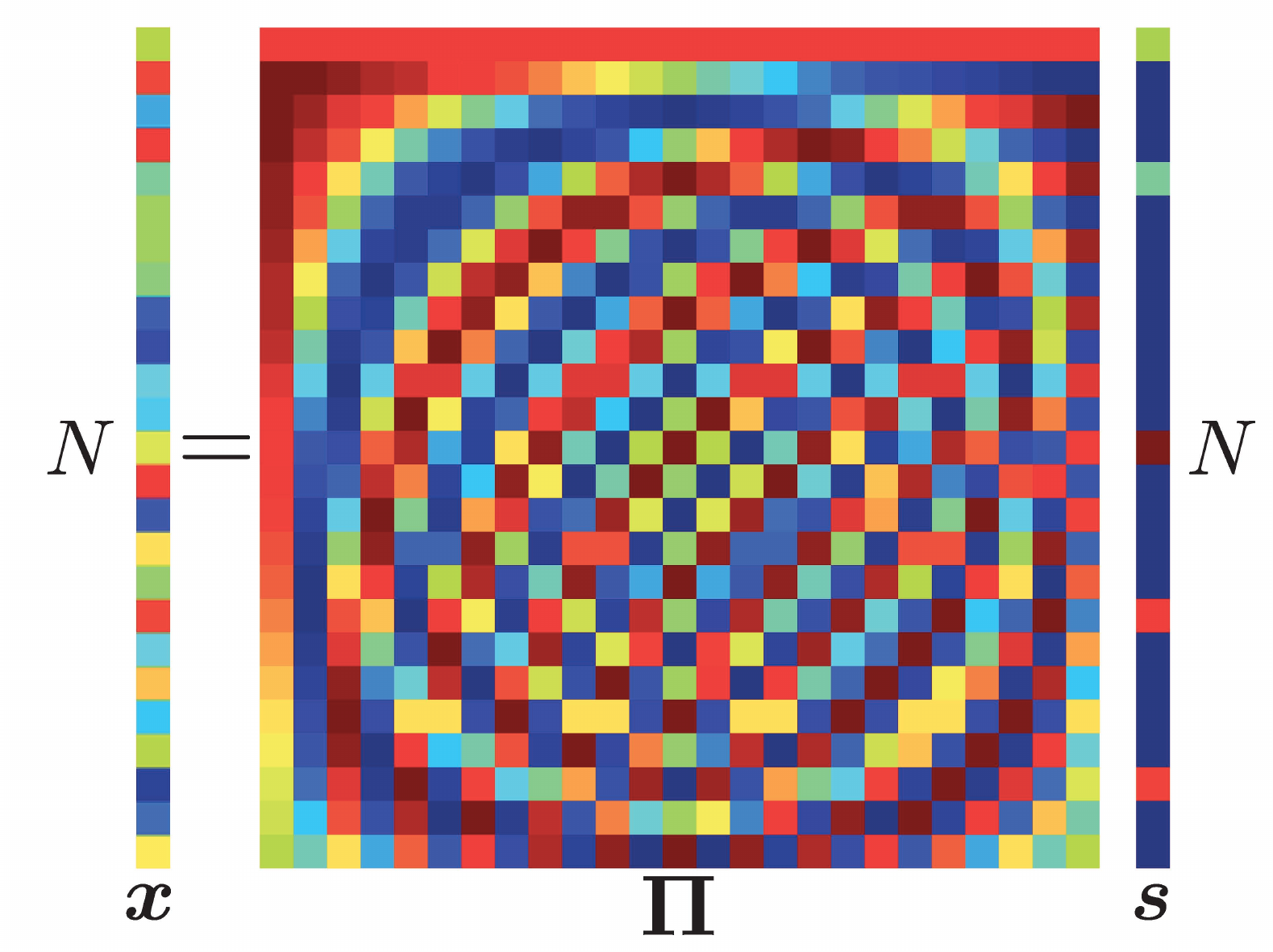}}
\caption{Comparison of various sensing schemes proposed in the
  literature (the noise term $\vomega$ is omitted for simplicity). We
  consider a discretized version of the sampled physical field that is
  contained into a vector $\vx$. In (a) we depict the traditional
  approach where we measure the physical field in each spatio-temporal
  location, thus having an identity operator $\mI$. In (b), we reduce
  the number of samples by taking random projections of the
  measurements. Note that we need to measure all the elements of $\vx$
  and we are just reducing the number of stored samples. On the other
  hand, in (c) we are reducing the number of measured samples using a
  sparse sampling matrix $\mPhi$. Note that the methods in (b) and (c)
  require a set of conditions regarding $\vx$ and $\mPhi$ to be
  satisfied \cite{Candes:2008pi}. Among these conditions, we note that
  $\vx$ must be sparse under a certain known dictionary $\mat{\Pi}$,
  see (d). }
\label{fig:intro_all}
\end{figure*}
\begin{figure}[!t]
\centering\includegraphics[height=0.2\newhcol]{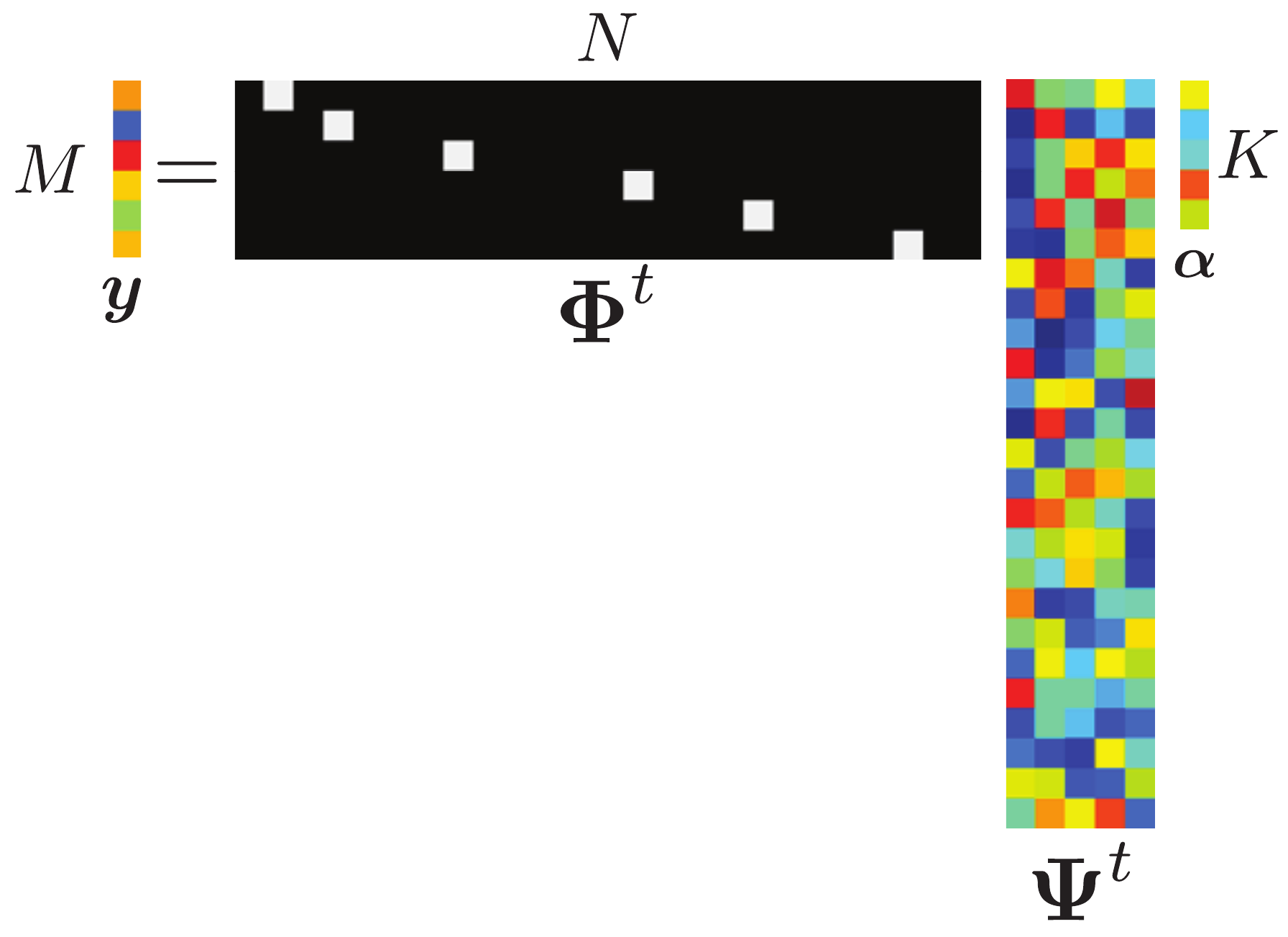}

\caption{ Graphical representation of the mathematical model of the
  proposed sensing scheme. The signal is modeled by an unknown
  time-varying linear $K$-dimensional model $\mPsi^t$ that is learn from
  the collected measurements. The sampling pattern $\mPhi^t$ is
  optimized at run-time according to the signal model and measures
  only $M$ values out of the $N$ available ones.}
\label{fig:intro_ss}
\end{figure}

One of the primary goals in designing a WSN is the reduction of the
energy consumption, to extend its lifetime without replacing or
recharging the batteries of sensor nodes.  The energy consumption of a sensor
node mainly comes from three activities: sensing, data-processing and
communication. Traditionally, the costs for processing and communication
are assumed to dominate the overall energy consumption, while the
cost for sensing is considered negligible. Therefore, a traditional WSN
collects as much data as possible, that is subsequently compressed and
transmitted with the lowest possible rate. In other words, it
collects a vector of samples $\vy_0$ that is equal to the discretized
physical field $\vx$ with some additive noise,
\begin{align}
\vy_0=\mI\vx + \boldsymbol{\omega},
\label{eq:trad_sensing}
\end{align}
where $\mI$ is the identity matrix of size $N$ and $\boldsymbol{\omega}$ represents
the noise; see Figure \ref{fig:eye} for an example.

If the energy consumed for sensing is comparable to that for
communication and data processing, ignoring the energy cost of the
former is sub-optimal. In fact, new sampling paradigms optimizing the
overall energy consumption emerge and show that further
reductions of the energy consumption are possible. The basic idea
involves a reduction of the number of collected samples and a
reconstruction of the missing data using algorithms exploiting the
structure available in the measured data. The reduction of the
collected samples is done by designing a sampling operator
$\mPhi\in\R^{M\times N}$ with $M\ll N$, that it is used instead of the
identity matrix as,
\begin{align}
\vy=\mPhi\vx + \boldsymbol{\omega}. \nonumber
\end{align}
Note that $\vy$ is significantly shorter than $\vx$ and the
reconstruction algorithm must estimate a significant amount of
information from a limited amount of data. Therefore,
regularization and constraints are added to the problem so that a
stable solution can be obtained. Moreover, the reconstruction
algorithm must be jointly designed with the sampling matrix $\mPhi$ to
obtain a precise estimate of $\vx$.

Pioneering work on sparse sampling considered compressive sensing
(CS) as a reconstruction scheme. CS attempts to recover $\vx$ by
solving a convex optimization problem, under the assumption that $\vx$
is sparse in a known dictionary $\mat{\Pi}$. However, the solution is
only approximate and it is exact if $\mat{\Pi}$ and $\mPhi$ satisfy
certain requirements that are generally hard to check
\cite{Candes2006}. Initially, \cite{Duarte2006, Wang2007a,Luo2009}
proposed the use of a sampling matrix $\boldsymbol\Phi$ composed of
random i.i.d. Gaussian entries. Note from Figure \ref{fig:dense} that
such $\mPhi$ has very few zero elements. Therefore, the number of
sensing operations is not actually reduced because we need to know all
the values of $\vx$ to compute $\vy$. Moreover, if we adopt a
distributed algorithm, a dense $\mPhi$ requires the sensor nodes to
transmit their local samples to the other nodes, causing an excessive
energy consumption for communications.

To overcome such limitations, \cite{Wu2012,Quer2012} proposed to use a sparse
matrix $\mPhi$ which contains very few non-zero
elements. More precisely, $\mPhi$ has generally only one non-zero
element per row and the locations of such elements determine the
spatio-temporal sampling pattern, see Figure
\ref{fig:sparse}. However, the sampling patterns in these schemes are
either fixed or randomly generated and thus not well adapted to the
measured signal.  Moreover, it is generally hard to guarantee the
recovery of a faithful representation of $\vx$, because the sparsity of
dictionary $\mat{\Pi}$ usually changes over time and it may not
satisfy the theoretical requirements of CS \cite{Candes:2008pi}.

Since the statistics of $\vx$ are often unknown and varying over time,
it may be advantageous to consider the decomposition
\begin{align}
  \vx=\mPsi^t \valpha,
\end{align}
where $\mPsi^t$ is the time-varying model and $\valpha\in\R^K$ is a
low dimensional representation of $\vx$ with $K\ll N$. While the
ignorance and the non-stationarity of the model $\mPsi^t$ forces us to
learn it from the samples collected in the past, it may give us the
advantage of optimizing the sampling pattern $\mPhi^t$ according to
$\mPsi^t$. Note that $\mPhi^t$ is also time-varying
as compared to the fixed pattern $\mPhi$ in Figure~\ref{fig:intro_all}.

This new problem statement raises new challenges. While the model
$\mPsi^t$ can be learnt from the incomplete measurements $\vy$ with
some effort using an online version of the principal component
analysis (PCA), the sampling scheduling problem is generally
combinatorial and hard to optimize. In this paper, we propose to
generalize FrameSense, an algorithm that generates a near-optimal
sensor placement for inverse problems \cite{Ranieri:2013wp}. Instead
of optimizing the sensor placement, we optimize the spatio-temporal
sampling pattern of the WSN. The obtained sampling pattern is
generally irregular, time-varying and optimized to gather the maximum
amount of information. In particular, it simultaneously exploits the
intra-node (temporal) and inter-node (spatial) correlation potentially
present in the data. See Figure \ref{fig:intro_ss} for a graphical
illustration of the low-dimensional model and of the irregular
sampling patterns. Note that the proposed method deviates from the
recent sparse sensing schemes~\cite{Quer2012,Wu2012} because the
sampling pattern is neither fixed nor random but dynamically adapted
to the signal's low-dimensional model.

It is worth mentioning that the proposed method imposes no on-sensor
computation nor inter-node communication. Each sensor node simply
collects measurements according to a designated sampling pattern and
transmits the data to a common server. The server receives all the
data from one or multiple sensor nodes and performs signal
reconstruction. This is actually in accordance to the setup of
distributed source coding~\cite{Viswanatha2012}, where no inter-node
communication is used. Hence, the proposed algorithm provides an
alternative solution to the distributed coding problem: the
communication rate is reduced and the reconstruction error is bounded
without using any inter-node communication.

The proposed algorithm is tested on different sets of real-word data,
outperforming both the traditional sensing schemes and the state-of-the-art
sparse sensing schemes, in terms of reconstruction quality of $\vx$
given a fixed amount of measurements. Given the aforementioned
characteristics, we call the proposed method ``\emph{Distributed Adaptive Sparse Sensing}'', or \emph{DASS}.

\section{Problem Formulation}
\label{sec2}
\begin{figure}[!t]
\centering
\includegraphics[height=0.19\newhcol]{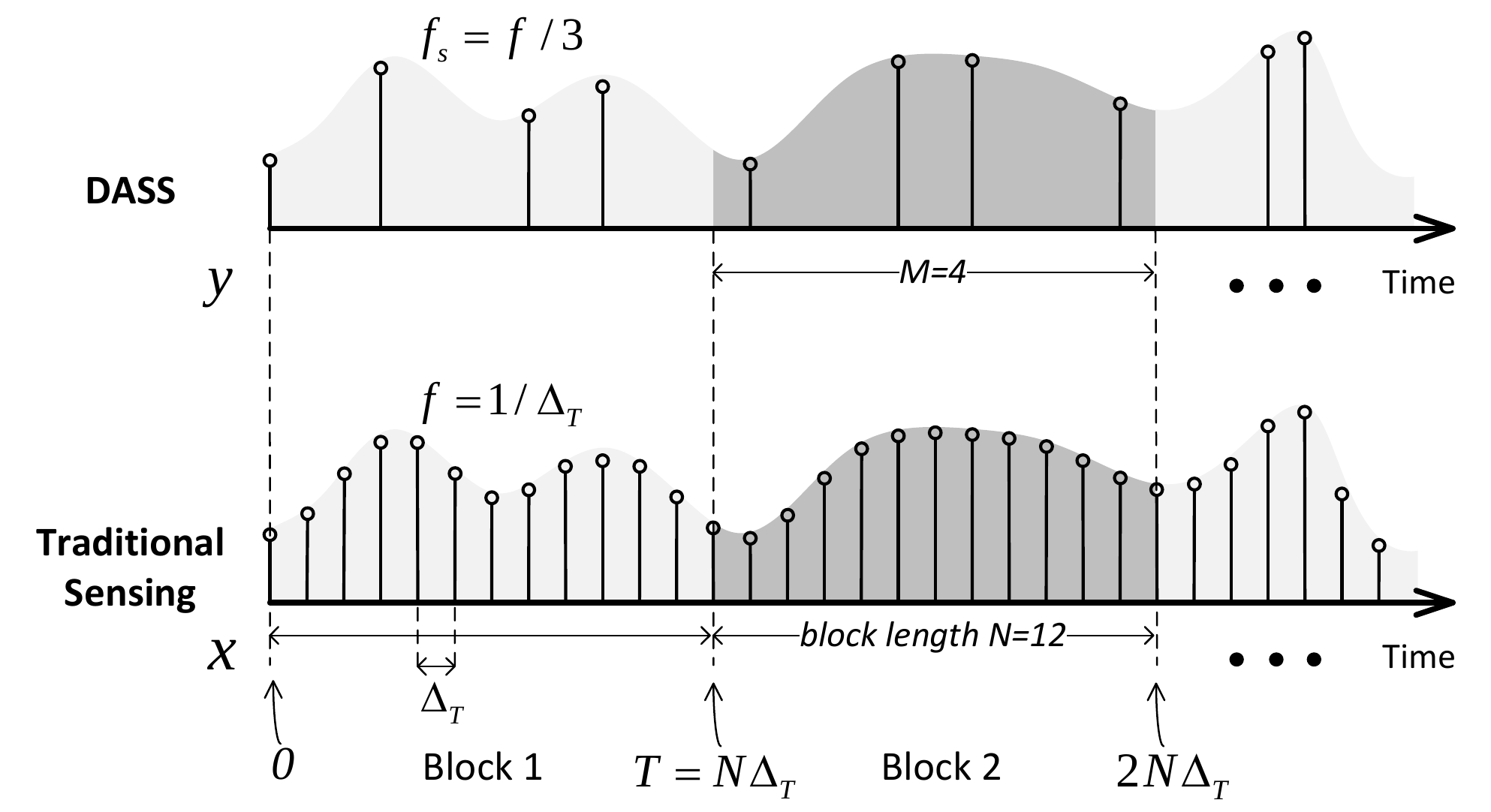}
\caption{Upper plot: optimized temporal sampling pattern of DASS. Lower plot: traditional sensing scheme, where
  samples are collected regularly in time. The subsampling factor is
  $\gamma=1/3$, since we collect 4 samples instead of 12 in each block.}
\label{sampling}
\end{figure}

In this section, we first state the sampling scheduling problem for a
WSN having just one sensor. At the end of the section, we generalize
the problem statement to a WSN with multiple nodes.  We consider a
block-based sensing strategy, meaning that the WSN samples the field
for a certain time $T$ and at the end we reconstruct the vector $\vx$
from the collected samples.  Note that the block length is known and
defined a-priori.

For each temporal block, the discrete physical field $\vx$ is composed
of $N$ samples of $x(\vp,t)$,
\begin{align}
  \vx=\left[x(\vp,0),x(\vp,\Delta_T),\cdots,x(\vp,(N-1)\Delta_T)\right]^\top,
\end{align}
where $\vp$ indicates the sensor node location and $\Delta_T$ is the
sampling period. Note that $\Delta_T$ determines the desired temporal
resolution and its inverse is the sampling frequency,
$f=1/\Delta_T$. The temporal duration of a block is $T=N\Delta_T$,
that is also the maximum delay this sensing scheme occurs---the larger
$T$, the longer the delay. See Figure \ref{sampling} for a graphical
representation of the physical field and its discrete version $\vx$.

We denote the reconstructed physical field obtained from the WSN
samples as $\widetilde \vx$. In a sparse sampling scenario, we aim at
reconstructing $\widetilde{\vx}$ from just a subset of elements of
$\vx$. More precisely, we measure $M$ elements out of $N$, where
$M<N$. The set of indices $\boldsymbol\tau^t=\{\tau^t_i\}_{i=1}^M$
denotes the indices of these $M$ samples and it is chosen adaptively
according to the previous measurements. Note that the sampling pattern
$\boldsymbol\tau^t$ uniquely determines the sampling matrix
$\mPhi^t\in\R^{M\times N}$:
\begin{align}
\mPhi^t_{i,j}=
\begin{cases}
  1\quad\text{ if } j=\tau^t_i \\
  0 \quad \text{ otherwise}
\end{cases}.\nonumber
\end{align}
It is important to underline that $\mtau^t$ is time-varying and
potentially changes at every block to adapt to the signal model
$\mPsi^t$.  Figure~\ref{sampling} shows an example of sampling patterns where
$\boldsymbol\tau^t$ changes for each block.

We define $f_s=\frac{M}{N}\cdot f=\gamma f$ to be the average sampling
frequency of the sensor node\footnote{Note that it is an average
  frequency given the irregular and time-varying sampling
  pattern.}. The subsampling rate $\gamma = f_s/f<1$ is an important
figure of merit for a sparse sampling algorithm---the lower the $\gamma$,
the lower the energy consumed for sensing.

The measured signal $\vy\in\R^{M}$ is defined as
\begin{align}
  \vy=\mPhi^t\vx +\vomega,
\end{align}
where $\vomega$ represents the measurement noise, that is modeled as an additive
white Gaussian noise (AWGN) with variance $\sigma^2$. Note that it is
reasonable to model the noise phenomena as AWGN since the thermal
effects~\cite{Johnson1928} or/and quantization~\cite{Widrow2008} are
often the dominating terms\footnote{Other noise model may be of
interest for specific sensors; for example the noise term of a Geiger
counter is usually modeled as a Poisson process.}.

The target of DASS is to optimize the sampling pattern $\mPhi^t$
at the $t$-th block according to $\mPsi^t$ such that we collect the
minimum number of samples $M$ while still being able to recover
precisely the original signal. Since we modeled the noise as a AWGN,
we assess the quality of the recovered signal by using root-mean-square
error (RMSE):
\begin{align}
\epsilon=\frac{1}{\sqrt{N}}\|{\vx}-\widetilde{\vx}\|_2. \nonumber
\end{align}

\begin{figure}[!t]
\centering
\includegraphics[height=0.19\newhcol]{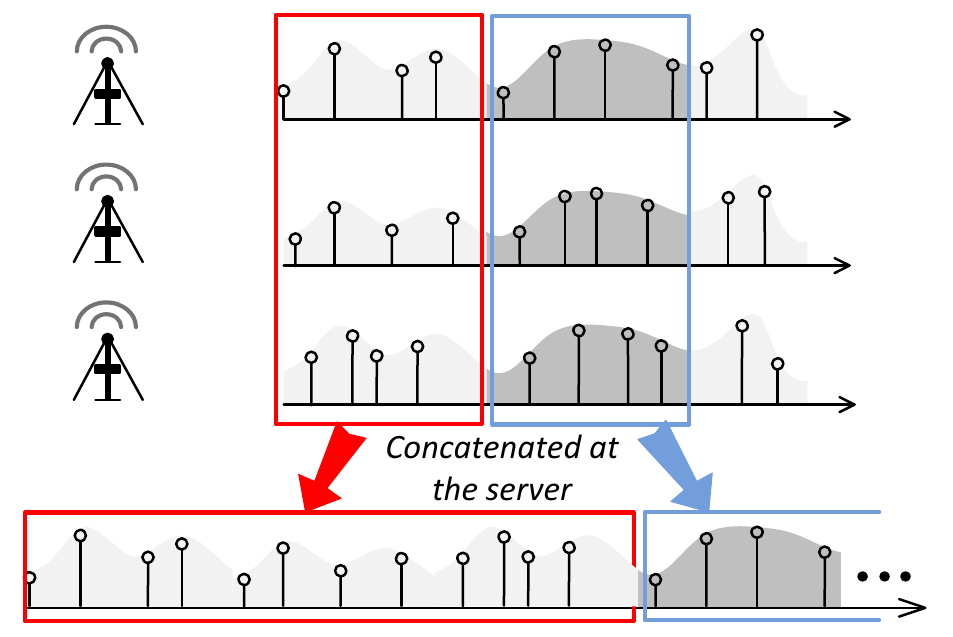}
\caption{Signals of multiple distributed sensor nodes can be concatenated into a single signal stream at the server for recovery.}
\label{app3}
\end{figure}

\begin{table}[!t]
\renewcommand{\arraystretch}{1.15}
\caption{Summary of notation}
\label{notation}
\centering
{\footnotesize\begin{tabular}{m{0.12in}m{1.25in}|m{0.12in}m{1.2in}}\hline
    $N$ &  desired number of samples in a block  & $M$ & number of measurements in a block, equals $\lfloor N\gamma\rfloor$\\
    $\Delta_T$ & temporal resolution of original signal  & $f$ &  sampling frequency of original signal, equals $1/\Delta_T$ \\
    $f_s$ &  average sampling frequency of the sensor  & $\gamma$ & subsampling rate $f_s/f$\\
    $\widetilde{\vx}$ &  reconstructed signal $\in\R^{N}$ & $\vx$ &  original signal $\in\R^{N}$ \\
    $\vy$ &  measured signal $\in\R^{M}$ & $\vomega$ &  measurement noise \\
    $\boldsymbol\tau^t$ & sampling pattern of the $t$-th block & $\boldsymbol\Phi^t$ & sampling matrix of the $t$-th block $\in\R^{M\times N}$ \\
    $\overline{\vx}$ & mean of the signal $\in\R^{N}$ & $\boldsymbol\Psi^t$ & signal model of the $t$-th block $\in\R^{N\times K}$ \\
    $\boldsymbol\alpha$ &  low dimensional representation of $\vx\in\R^{K}$ & $\widetilde{\boldsymbol\Psi}^t$ & rows of $\boldsymbol\Psi^t$ selected by $\boldsymbol\tau^t$ $\in\R^{M\times K}$ \\
\hline
\end{tabular}}
\end{table}

\emph{Multiple-node scenario}: while the above problem statement
focuses on a single-sensor scenario for simplicity of notation, it is
simple to generalize the statement to a WSN with more than one sensor
node. More precisely, we assume that the nodes are synchronized, so that we can
concatenate all the measured blocks at different locations $\vp_i$ in
a unique signal block $\vx$, see Figure \ref{app3} for an example. The
sparse sampling problem is generalized to a spatio-temporal domain
meaning that we have to choose \emph{when and where} we want to sample
to collect the maximum amount of information.

\section{Building Blocks}
\label{sec3}
\begin{figure}[!t]
\centering
\includegraphics[height=0.13\newhcol]{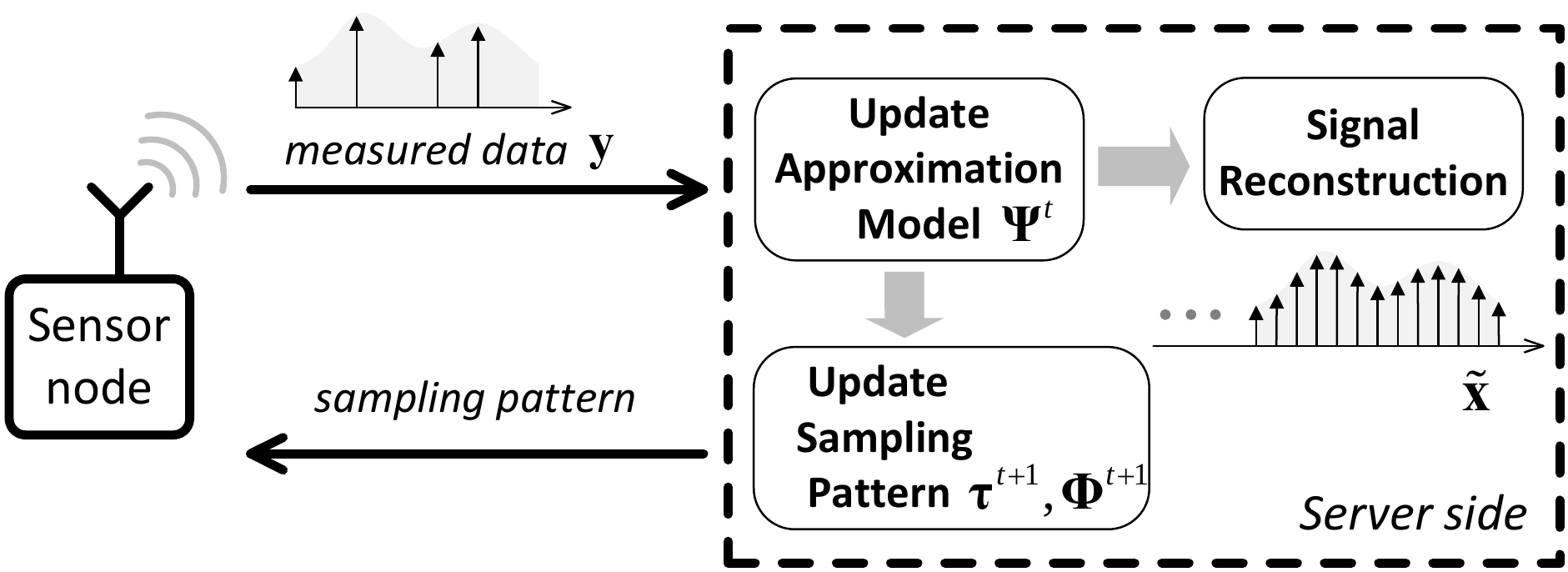}
\caption{Representation of the operations of DASS in a WSN. The sensor
  node sends the measured data to the processing server and
  receives the sampling pattern for the next temporal block. The
  server uses the data to update the signal model $\mPsi^t$,
  reconstructs the discrete physical field $\widetilde{\vx}$ and
  optimizes the sampling pattern $\mtau^{t+1}$ for the sensor nodes.
  Note that $\mtau^{t+1}$ uniquely determines $\mPhi^{t+1}$.}
\label{framework}
\end{figure}

The proposed method is graphically represented in
Figure~\ref{framework} and is based on the three building blocks
described in this section:

\begin{compactenum}[1)]
\item The desired signal $\widetilde{\vx}$ is reconstructed using the
  collected measurements $\vy$, the signal model $\mPsi^t$
  and the estimated mean $\overline{\vx}$ (Section~\ref{sec3.1}).
\item The measurements $\vy$ are used to update the approximation
  model $\mPsi^t,\overline{\vx}$ (Section~\ref{sec3.2}).
\item The sampling pattern for the next temporal block $\mtau^{t+1}$
  is optimized according to $\mPsi^t$ and is transmitted back to the
  sensor node(s) (Section~\ref{sec3.3}).
\end{compactenum}

The overhead of DASS on the sensor node is minimal in practice.
First, the sampling pattern $\mtau^{t}$ has a sparse structure and hence
it can be encoded efficiently with a few bytes per block. Therefore, the
extra communication cost for receiving $\mtau^{t}$ is minimal.
Second, all the algorithmic complexity of DASS is at the
server side, while the sensor nodes only need to sample and transmit
the signal according to the sampling pattern received from the
server. Therefore, the CPU and memory requirements of the sensor node
are minimal.

In what follows, we analyze each block explaining the challenges and
the proposed solution.

\subsection{Signal Approximation and Reconstruction}
\label{sec3.1}

Due to the nature of most physical fields, a signal block is partially
predictable by analyzing past data. In many cases, this predictability
can be expressed by assuming that the signal belongs to a
$K$-dimensional linear subspace $\mPsi^t\in\R^{N\times K}$. Such a
subspace approximates $\vx$ as
\begin{align}
\widehat \vx=\mPsi^t\valpha +\overline{\vx},
\label{eq5.1.1}
\end{align}
where $\widehat{\vx}$ is the approximated field, $\valpha\in\R^{K}$
is the vector of the projection coefficients and $\overline{\vx}$ is the mean of
$\vx$.

If the modeling subspace $\mPsi^t$ is well designed and $K$ is
sufficiently large compared to the complexity of $\vx$, the signal
realization $\vx$ can be accurately expressed with just $K<< N$
coefficients contained in $\valpha$. To find such a subspace, we analyze
all the past signal realizations and estimate at the $t$-th block the
$K$-dimensional subspace $\mPsi^t$ that minimizes the expected
approximation error
\begin{align}
\epsilon_a=\frac{1}{\sqrt{N}}\mathbb{E}\left(\|\vx-\widehat\vx\|_2\right). \nonumber
\end{align}
This is a dimensionality reduction problem that can be solved by the
well known technique of \textit{principal component analysis
  (PCA)}. It has an analytic solution but it requires the covariance
matrix $\mC_{\vx}$.

Unfortunately, in our scenario it is hard to estimate $\mC_{\vx}$
since we have access only to $M$ out of $N$ elements of
$\vx$. However, if the $M$ sampled elements are varying at each
temporal block $t$, we may collect enough information to have a
sufficiently precise estimate of $\mC_{\vx}$. We present a set of
methods to estimate $\mC_{\vx}$ in Section~\ref{sec3.2}.

Note that the approximation through $\mPsi^t$ exploits the correlation
among the elements of $\vx$. The higher the correlation available in
$\vx$, the lower the dimensionality of the subspace $\mPsi^t$, the
number of parameters $K$ and the necessary measurements $M$. Hence,
one of the key aspects is the choice of the signal block length
$T$. In fact, it should be chosen such that the delay of the WSN
respects the design specification while maximizing the correlation
among the blocks. For example, if we consider a sensor measuring the
outdoor light intensity, the signal itself naturally has diurnal patterns.
If we choose a block length of one hour, the correlation between
the signal block is usually weak. On the other hand, if we choose
a block length of one day, the correlation is stronger due to the
aforementioned patterns.

Once the approximation model $\mPsi^t$ is estimated, the task of
recovering the signal $\widetilde{\vx}$ amounts to estimating $\valpha$
from the measurements $\vy$ when considering the approximated signal
model
\begin{align}
  \vy\approx\mPhi^t\widehat{\vx}+\vomega=\mPhi^t(\mPsi^t\valpha+\overline{\vx})+\vomega. \label{eq:meas}
\end{align}
In general, we can recover $\valpha$ by solving an Ordinary Least Square
(OLS) problem:
\begin{equation}
\widetilde{\valpha} = {\arg\min_{\valpha}}\|\vy-\mPhi^t\overline{\vx}-\mPhi^t\mPsi^t\valpha\|_2^2,
\label{eq5.2.4}
\end{equation}
which has the following analytic solution
\begin{align}
\widetilde{\valpha}=(\mPhi^t\mPsi^t)^\dagger(\vy-\mPhi^t\overline{\vx}).
\end{align}
Here $(\mPhi^t\mPsi^t)^\dagger$ is the Moore-Penrose pseudoinverse of
$\mPhi^t\mPsi^t$ that is defined for a generic matrix $\mA$ as
$\mA^\dagger=(\mA^*\mA)^{-1}\mA^*$.

The reconstruction algorithm is straightforward and is described in
Algorithm~\ref{algoreconstruct}. The following theorem states the
necessary conditions to find a unique solution and provides an upper
bound for the reconstruction error, that is going to be fundamental
when optimizing the sampling pattern.

\begin{theorem}
  Consider a sensor network measuring a physical field as in
  \eqref{eq:meas} where the measurements are corrupted by an
  i.i.d. Gaussian noise with variance $\sigma^2$.  If $M\geq K$,
  $\mPsi^t$ is formed by orthonormal columns and
  $\operatorname{rank}(\mPhi^t\mPsi^t)=K$, then $\widetilde{\vx}$ can
  be uniquely determined using Algorithm~\ref{algoreconstruct}.  The
  reconstruction error is bounded by
\begin{equation}
\epsilon^2=\frac{1}{N}\|\vx-\widetilde{\vx}\|_2^2\leq\frac{1}{\lambda_K}\epsilon_a^2+\sigma^2\sum^K_{k=1}\frac{1}{\lambda_k},
\label{eq5.2.6}
\end{equation}
where $\epsilon_a$ is the approximation error due to the signal model
$\mPsi^t$ and $\lambda_i$ are the eigenvalues of
${\mPsi^t}^*{\mPhi^t}^*\mPhi^t\mPsi^t$ sorted in decreasing order.
\label{theorem1}
\end{theorem}

\begin{IEEEproof}
Since the Gaussian noise is independent from the approximation error,
we can treat them independently. Moreover, it is sufficient to compute
the error on the estimation of $\valpha$ given the orthonormality
of the columns of $\mPsi^t$.

For the approximation error $\epsilon_a$, we look at the worst case
scenario with the following optimization problem
\begin{align}
&\max \quad \|(\mPsi^t\mPsi^t)^\dagger(\vx-\widehat{\vx})\|^2_2 \nonumber\\
&\text{subject to}\quad
\frac{1}{N}\|(\vx-\widehat{\vx})\|^2_2=\epsilon_a \nonumber,
\end{align}
whose solution is proportional to the largest eigenvalue of
$(\mPsi^t\mPsi^t)^\dagger$. More precisely, it is possible to show
that approximation noise is equal to the
$\frac{1}{\lambda_K}\epsilon_a^2$, where $\epsilon_a$ is the norm of
the approximation error.

For the white noise, we consider a previous result given in
\cite{Fickus:2011vq} to conclude the proof.
\end{IEEEproof}

\begin{algorithm}[!t]
\caption{Signal reconstruction}
\label{algoreconstruct}
{
\begin{algorithmic}[1]
\Require $\mPsi^t$, $\overline{\vx}$, $\mtau^t$ and $\mPhi^t$
\Ensure $\widetilde{\vx}$
\State Measure the signal $\vy$ according to $\mtau^t$.
\State $\widetilde{\vx}=\mPsi^t(\mPhi^t\mPsi^t)^\dagger(\vy-\mPhi^t\overline{\vx})+\overline{\vx}$.
\end{algorithmic}}
\end{algorithm}

The upper-bound of the total error $\epsilon$ is a
function of both the approximation error $\epsilon_a$ and measurement
noise. The former term depends on the number of parameters $K$: when
$K=N$, we have $\epsilon_a=0$ and it grows when we decrease $K$. However, the rate at which the error increases depends on the
spectrum of $C_{\vx}$. In fact, if $\vx$ has elements that are highly
correlated, a small $K$ could be sufficient to model $\vx$ with a
small approximation error. The latter term can be controlled directly
by optimizing the sampling pattern. More precisely, we
cannot reduce $\sigma$ but we can reduce the amplification due to the
spectrum $\lambda_k$ through an optimization of the sampling matrix
$\mPhi^t$.

Note that the part involving $\epsilon_a$ depends only on the smallest
eigenvalue because we are not guaranteed that the approximation error
\emph{spreads} over all the eigenvectors of $\mPhi^t\mPsi^t$. In fact,
the worst case scenario is represented by the approximation error
being in the same direction of the eigenvector with the smallest
eigenvalue and $\epsilon_a$ is consequently maximally amplified.

Compared to the methods based on CS, our approach based on a
low-dimensional model and OLS has the following advantages: i) the solution
is easy to compute and it requires a single matrix inversion, ii) it
enables an analysis of the reconstruction error and a consequent
optimization of the sampling pattern $\mtau^t$ such that $\epsilon$ is
minimized.

\subsection{Learning from Incomplete Data Over Time}
\label{sec3.2}

In Section \ref{sec3.1}, we have highlighted some challenges regarding
the estimation of the covariance matrix $\mC_{\vx}$ --- a fundamental
step to determine the approximation model $\mPsi^t$. Most of the
challenges derive from the lack of a sufficiently large set of
realizations of $\vx$, that are needed to estimate $\mC_{\vx}$. First,
there is virtually no past data for a newly installed WSN. Second,
$\mC_{\vx}$ is likely to vary over time. Third, a high ratio of data
points ($1-\gamma$) are not available for the estimation since we
collect sparse measurements. Therefore, we need an on-line algorithm
that estimates and adaptively updates the covariance matrix
$\mC_{\vx}$ from incomplete data.

\begin{algorithm}[!t]
\caption{Updating $\mPsi^t,\overline{\vx}$ using a buffer}
\label{updater1}
{
\begin{algorithmic}[1]
\Require $\vy$, $L$
\Ensure $\mPsi^t,\overline{\vx}$
\State interpolate $\vy\to \vx_{\textrm{intep}}$.
\State insert $\vx_{\textrm{intep}}$ into a buffer storing the most recent $L$ blocks.
\State estimate $\mathbf{C}_{\vx}$ and $\overline{\vx}$ from the buffer.
\State $\mPsi^t$ is formed by the first $K$ eigenvectors of $\mathbf{C}_{\vx}$ ordered in decreasing values of its eigenvalues.
\end{algorithmic}}
\end{algorithm}

\begin{algorithm}[!t]
\caption{Updating $\mPsi^t,\overline{\vx}$ using incremental PCA}
\label{updater2}
{
\begin{algorithmic}[1]
\Require $\vy$, $L$, $\mPsi^{t-1}$, $\boldsymbol\lambda^{t-1}, \overline{\vx}^{t-1}$
\Ensure $\mPsi^t, \boldsymbol\lambda^t, \overline{\vx}^t$
\State interpolate $\vy\to \vx_{\textrm{intep}}$.
\State $\mathbf{a}={\mPsi^{t-1}}^*(\vx_{\textrm{intep}}-\overline{\vx}^{t-1})$.
\State $\mathbf{b}=\left(\mPsi^{t-1}\mathbf{a}+\overline{\vx}^{t-1})\right)-\vx_{\textrm{intep}}$, and then normalize $\mathbf{b}$.
\State $c=\mathbf{b}^*(\vx_{\textrm{intep}}-\overline{\vx}^{t-1})$.
\State $\mathbf{D}=\frac{1}{L+1}\left[
           \begin{array}{cc}
             \textrm{diag}(\boldsymbol\lambda^{t-1}) & \boldsymbol 0 \\
             \boldsymbol 0^* & 0 \\
           \end{array}
         \right]+\frac{L}{(L+1)^2}\left[
           \begin{array}{cc}
             \mathbf{a}\mathbf{a}^* & c\mathbf{a} \\
             c\mathbf{a}^* & c^2 \\
           \end{array}
         \right]$.
\State Solve the eigenproblem: $\mathbf{D}=\mathbf{R}\cdot\textrm{diag}(\boldsymbol\lambda')\cdot\mathbf{R}^{-1}$, $\boldsymbol\lambda'$ is sorted in decreasing order.
\State $\boldsymbol\Psi'=\left[\mPsi^{t-1}\ \mathbf{b}\right]\cdot \mathbf{R}$.
\State update $\mPsi^t$ as the first $K$ columns of $\boldsymbol\Psi'$.
\State update $\boldsymbol\lambda^t$ as the first $K$ values of $\boldsymbol\lambda'$.
\State update $\overline{\vx}^t$ as $\left(L\overline{\vx}^{t-1}+\vx_{\textrm{intep}}\right)/(L+1)$.
\end{algorithmic}}
\end{algorithm}

The main difficulty is the lack of complete
realizations of $\vx$. Two strategies are generally considered to
overcome such a problem. The first one proposes to estimate from $\vy$
an interpolation $\vx_{\text{interp}}$ using classic interpolation
methods such as linear, polynomial or spline interpolation. The second
strategy skips the estimation of $\mC_{\vx}$ and attempts to perform
directly the principal component analysis with the data having missing
entries, see \cite{Raiko2008}.

In our experience, the second class of algorithms is less performant
for our purposes. Therefore, we focus our attention on the
interpolation methods.  More precisely, we analyze two different
methods that implement an adaptive learning and updating of the
approximation model $\mPsi^t$ from the interpolated signal
$\vx_{\textrm{intep}}$: Algorithm~\ref{updater1} and
Algorithm~\ref{updater2}.

Algorithm~\ref{updater1} uses a FIFO buffer to store the most recent
$L$ blocks. Whenever a new block is added into the buffer, the oldest
block in the buffer is excluded. As the approximation model is
estimated according to the signal realizations in the buffer, this
scheme is able to capture the variation of signal statistics over
time.

Algorithm~\ref{updater2} adaptively updates the approximation model
via a technique called incremental PCA~\cite{Hall1998}. It does not
keep signal realizations in memory, instead, it stores the largest
$K$ eigenvalues of $\mC_{\vx}$, $\boldsymbol\lambda=\{\lambda_i\}$,
for $i=1,\cdots,K$. This method requires significantly less memory
($K$ versus $N\times L$), and shows better performance when compared
to Algorithm~\ref{updater1}. Note that in both algorithms, the choice
of $L$ depends on the variability of the signal statistics for each
specific application. In practice, we can cross-validate this
parameter to find a suitable value (e.g., $L=30$). We discuss and
compare the performance of these two algorithms in the experimental results.

\subsection{Sampling Scheduling Algorithm}
\label{sec3.3}
According to Theorem \ref{theorem1}, minimizing the overall error
$\epsilon$ is equivalent to finding the optimal sampling pattern
$\boldsymbol\tau$ that minimizes (\ref{eq5.2.6}).
In this paper, we assume that the model $\mPsi^t$ is
sufficiently precise and the dimensions $K$ is large enough
so that the term due to the white noise $\sigma$ is dominant.

Therefore, we would like to find the sampling pattern
that minimizes the following cost function,
\begin{align}
\Theta(\widetilde{\mPsi}^t)=\sum_{k=1}^K\frac{1}{\lambda_k},
\end{align}
where $\lambda_k$ are the eigenvalues of
$(\widetilde{\mPsi}^t)^*\widetilde{\mPsi}^t$, and
$\widetilde{\mPsi}^t=\mPhi^t\mPsi^t.$ Note that this optimization is
equivalent to finding the $M$ rows of $\mPsi^t$ that forms the
submatrix $\widetilde{\mPsi}^t$ with the smallest
$\Theta(\widetilde{\mPsi}^t)$. However, it has been already shown that
such optimization is NP-hard \cite{Das:2008uc} and has a complexity
$\mathcal{O}\left(\binom{N}{M}\right)$, which is prohibitively high in
practice.

In this section, we investigate approximate solutions to the
scheduling problem that can be implemented efficiently. These
approximate solutions are usually hard to find because the cost
function $\Theta(\widetilde{\mPsi}^t)$ has many local minima that are
arbitrarily far away from the global minimum. Therefore, proxies of
$\Theta(\widetilde{\mPsi})$ are usually chosen as a cost function for
the approximated algorithm with a twofold aim: (i) inducing an indirect
minimization of $\Theta(\widetilde{\mPsi}^t)$ and (ii) being
efficiently optimized by standard techniques, as convex optimization
or greedy algorithms.

In this paper, we extend our recent work \cite{Ranieri:2013wp} about
optimal sensor placement to solve the sampling scheduling
problem. In fact, if we define the linear inverse problem to be the
estimation of $\vx$ from $\vy$, then the sensor scheduling problem is
equivalent to sensor placement.  The algorithm \cite{Ranieri:2013wp}
optimizes the sensor placement by a greedy minimization of the frame
potential \cite{Casazza:2006wl}, that is defined as
\begin{align}
  \operatorname{FP}(\mPsi^t,\calS)=\sum_{i,j\in\calS}|\langle\mpsi_i,\mpsi_j\rangle|^2,
\label{eq:cost_function}
\end{align}
where $\mpsi_i$ is the $i$-th row of $\mPsi^t$ and $\mathcal{S}$
contains the set of candidate locations for sensing. Under some mild
solutions, we proved that such an algorithm is near-optimal
w.r.t. the RMSE of the solution.

In this work, we propose a sampling scheduling algorithm based on an
equivalent greedy ``worst-out'' procedure: as input we have the signal
model $\mPsi^t$ and we initially consider the identity matrix of size
$N$ as the sampling matrix $\mPhi^{t+1}$. At each iteration, we remove
the row of $\mPhi^{t+1}$ that maximizes
\eqref{eq:cost_function}. After $N-M+1$ iterations we are left with an
optimized $\mPhi^{t+1}$ that has only $M$ elements different from zero
and has near-optimal performance when reconstructing $\vx$ from the
measurements $\vy$. Note that if $\mPsi^t$ satisfies the conditions
given in \cite{Ranieri:2013wp}, the obtained sampling matrix
$\mPhi^{t+1}$ stably recovers $\vx$ from the measurements $\vy$.

Furthermore, since a uniform sampling schedule is a commonly-used
strategy that yields good performance in real
applications~\cite{Wu2012}, we compare it with the result returned by
the greedy algorithm and opt for the one with smaller reconstruction
error. Note that this error is approximated by the bound provided by Theorem~\ref{theorem1}.
A detailed description of the overall algorithm is given in Algorithm
\ref{alg:greedy}.

\begin{algorithm}[!t]
\caption{Greedy sampling scheduling}
\label{alg:greedy}
{
\begin{algorithmic}[1]
\Require $\boldsymbol{\Psi}^t,$ $M$
\Ensure $\boldsymbol\tau^{t+1}$ for the next temporal block
\State Initialize the set of removed sampling indices: $\mathcal{L}=\emptyset$.
\State Initialize the set of selected sampling indices: $\mathcal{S}=\{1,\cdots,N\}$.
\State Find the first two rows to eliminate, $\mathcal{L}=\arg \max_{i,j\in \mathcal{S}} |\left<\boldsymbol{\psi}_i, \boldsymbol{\psi}_j\right>|^2$.
\State Update $\mathcal{S}=\mathcal{S}\backslash\mathcal{L}$.
\Repeat
\State Find the optimal row, $i^*=\arg\max_{i\in \mathcal{S}}
 \operatorname{FP}(\mPsi^t,\mathcal{S}\backslash i)$.
\State Update the set of removed indices, $\mathcal{L}=\mathcal{L}\cup i^*$.
\State Update the set of selected indices, $\mathcal{S}=\mathcal{S}\backslash i^*$.
\Until{$|\mathcal{S}|=M$}
\State {\small$\boldsymbol\tau^{t+1} = {\arg\min_{\boldsymbol\tau}\,}\left \{ \frac{\epsilon_a^2}{\lambda_K}+\sigma^2\Theta(\widetilde{\boldsymbol\Psi}^t),\boldsymbol\tau \textrm{ is uniform pattern or } \mathcal{S}\right\}$.}
\end{algorithmic}}
\end{algorithm}

\section{Comparisons with Baseline Methods}
\label{sec4}
In this section, we briefly summarize the state-of-the-art methods for
the sparse sensing problem. They will serve as the baseline for
comparisons in Section~\ref{sec5}.

The first category of methods~\cite{Quer2012,Wu2012} are based
on compressive sensing (CS).  With the notations introduced in
Section~\ref{sec2}, $\vx$ is the unknown signal, $\vy$ contains the
incomplete measurements, and $\boldsymbol\Phi$ is a sparse sampling
matrix with only $M$ elements different from zero. We assume $\vx$ to
be sparse w.r.t. a dictionary $\boldsymbol\Pi$. More precisely, we
have $\vx= \boldsymbol\Pi \vs$ and $\vs$ has just a few coefficients
different from zero, that is $\|\vs\|_0\ll N$ (see~\cite{Cand`es2006}
for more details). By approximating the $\ell_0$ norm with the
$\ell_1$ norm~\cite{Candes2006}, the reconstruction method for the
noiseless case is:
\begin{equation}
\min_{\vs\in \mathbb{R}^N}\|\vs\|_1,\ \textrm{s.t.}\ \ \vy=\boldsymbol\Phi \boldsymbol\Pi \vs,
\label{eq4.1}
\end{equation}
while the one  for the noisy case is
\begin{equation}
\min_{\vs\in \mathbb{R}^N}\parallel\vs\parallel_1,\ \textrm{s.t.}\ \ \|\vy-\boldsymbol\Phi \boldsymbol\Pi \vs\|_2\leq \xi,
\label{eq4.2}
\end{equation}
where $\xi$ measures the energy of the noise.  Problem (\ref{eq4.1})
and (\ref{eq4.2}) are both convex and can be solved~\cite{Candes2006}
in polynomial time using various solvers, in general iterative or
based on convex optimization. In both methods, we use uniform sampling as the sampling scheduler ---
$\tau_j^t=\lfloor j N/M\rfloor$.

The second category of baseline methods~\cite{Quer2012} are based on learning the
$K$-dimensional time-varying model $\mPsi^t$ and a reconstruction via
OLS as in Algorithm~\ref{algoreconstruct}. We use two sampling
schedulers, namely, a uniform sampling, and a random sampling where $\tau_j^t$ is
randomly selected with a uniform distribution.

Table~\ref{baselines} lists all the methods (including DASS)
that are evaluated in the experiments.
To have a fair comparison, $\boldsymbol\Pi$ in CS-based methods and
${\boldsymbol\Psi}^t$ in OLS-based methods are both learnt\footnote{The experimental
  results show that $K=M$ is the best choice for CS-based methods,
  while $K<M$ is a parameter which needs to be optimized for OLS-based
  methods, see Section~\ref{sec5.1}.}
by the incremental PCA described in Algorithm~\ref{updater2}.

\begin{table}[!t]
\renewcommand{\arraystretch}{1.1}
\caption{Summary of methods used in experiments}
\label{baselines}
\centering
{\footnotesize\begin{tabular}{m{0.7in}|m{0.7in}|m{0.6in}}
\hline
Abbreviation & Reconstruction Algorithm & Sampling Scheduling \\
\hline
CS & (\ref{eq4.1}) & uniform \\
CSN & (\ref{eq4.2}) & uniform \\
OLS-random & Alg.~\ref{algoreconstruct} & random \\
OLS-uniform & Alg.~\ref{algoreconstruct} & uniform \\
DASS & Alg.~\ref{algoreconstruct} & Alg.~\ref{alg:greedy} \\\hline
\end{tabular}}
\end{table}

\section{Evaluations of DASS and Sparse Sensing Methods}
\label{sec5}
\begin{figure}[!t]
\centering
\includegraphics[height=0.16\newhcol]{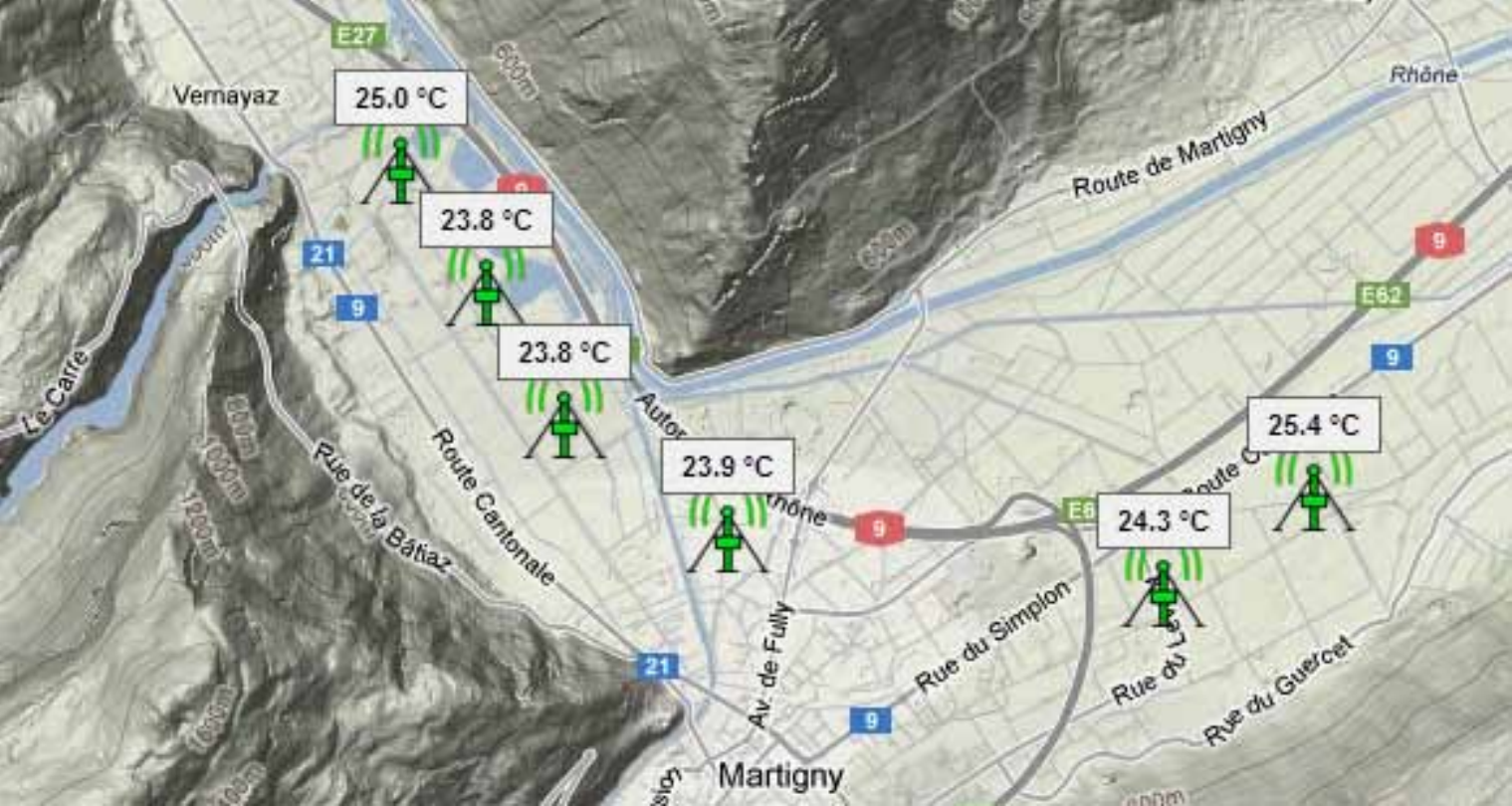}
\caption{Locations of the sensor nodes that collected the data-set \dataB{}.}
\label{map}
\end{figure}

In this section we evaluate the performance of DASS and compare it
with the state-of-the-art sparse sensing methods. Besides the
experiments on the single-node case, we also verify DASS in the
multi-node case where nearby sensor nodes measure spatially correlated
signals.  We use two real-world meteorological datasets as the ground
truth, namely \dataA{} and \dataB{}:
\begin{itemize}
\item \dataA{} is provided by \meteoswiss{}\meteoswissref{}. This dataset
contains 1500 days of continuous measurements for two physical
quantities (temperature and solar radiation)\footnote{We denote by
  \dataA{}-temperature the dataset of temperature measurements. The
  notation is similar for solar radiation.}, which are suitable for
studying long-term performance of DASS. As \meteoswiss{} only deployed
a few observation stations across the whole nation, we use \dataA{} for
evaluating the single-node case.
\item \dataB{} is provided by a microclimate monitoring service
provider~\sensorscope{}. A total of six stations are deployed in a
mountain valley (Figure~\ref{map}), covering an area of around
$18\ \textrm{km}^2$. The deployments were started in March 2012
and collected 125 days of continuous temperature measurements. We use
\dataB{} for evaluating the multi-node case.
\end{itemize}
The two datasets are
summarized in Table~\ref{dataspec}.
For both datasets, there are 144 uniformly sampled data points for
each day. We choose the day as the length of each block, that is, $N=
144$.

One of the targets of this section is to evaluate DASS and compare it
with other algorithms for different SNR regimes of the measurement. Since we cannot
measure directly the real value of the physical field, we assume that
\dataA{} and \dataB{} represent the real value of the field
$\vx$. Then, we add white gaussian noise to simulate the effect of
noisy measurements.

Note that the main merit figure considered in this section is the
final reconstruction error under a fixed subsampling rate
$\gamma$. Since all sparse sensing schemes directly transmit the
sensing samples without further data compression, two schemes with the
same $\gamma$ have the same amount of energy consumed for sensing and
communication\footnote{The processing costs of the considered sparse
  sensing methods are negligible.}, regardless of which sensing
platform is used.

\begin{table}[!t]
\renewcommand{\arraystretch}{1}
\caption{Summary of experimental datasets}
\label{dataspec}
\centering
{\footnotesize\begin{tabular}{m{0.5in}|m{1.0in}|m{0.6in}|m{0.5in}}
\hline
Dataset name & Physical quantity & Number of nodes & Number of days\\
\hline
\dataA{} & temperature, solar radiation & 1  & 1500 \\
\dataB{} & temperature &  6  & 125 \\\hline
\end{tabular}}
\end{table}

\subsection{Components of DASS}
\label{sec5.1}
In this section, we evaluate the key components of DASS, including the optimal choice of $K$,
the cost function $\Theta(\mPhi^t\mPsi^t)$ in the sampling scheduling algorithm,
and the performance of adaptive learning algorithms.

\begin{figure}[!t]
\centering
\includegraphics[height=0.135\newhcol]{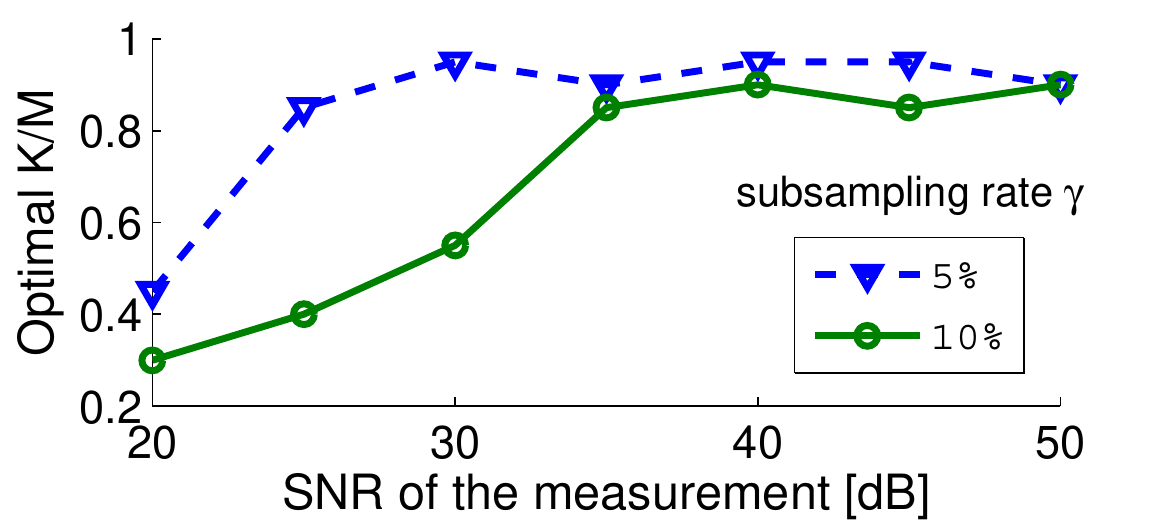}
\caption{Optimal ratio $K/M$ of DASS w.r.t. SNR of the measurement, for \dataA{}-temperature. Note $K/M$ must be smaller than 1 according to Theorem~\ref{theorem1}.}
\label{plot1}
\end{figure}

\keyitem{Optimal Choice of Dimension $K$} As stated in Theorem~\ref{theorem1},
the overall reconstruction error $\epsilon$ is a function of
both the approximation error $\epsilon_a$ and the cost function $\Theta(\mPhi^t\mPsi^t)$. Generally,
$\epsilon_a$ decreases with $K$ and $\Theta(\mPhi^t\mPsi^t)$ increases with $K$,
hence there is an optimal choice of $K$ for minimizing the overall
error. The optimal $K$ depends on the data statistics, the subsampling
rate, and the SNR of the measurement. By cross-validation, Figure~\ref{plot1} shows the
optimal ratio $K/M$ for \dataA{}-temperature. We can see that
DASS generally opts for a larger $K$ when the SNR of measurement increases.

\keyitem{Sampling Scheduling} The greedy algorithm proposed in
Section~\ref{sec3.3} (Algorithm~\ref{alg:greedy}) finds an approximate
solution of the sampling scheduling problem. By
Theorem~\ref{theorem1}, $\Theta(\mPhi^t\mPsi^t)$ determines the
reconstruction error. Table~\ref{condtab} shows the value of
$\Theta(\mPhi^t\mPsi^t)$ achieved by different sampling scheduling
methods for different datasets. Note that a higher value indicates
worse stability w.r.t. noise. We can see that the greedy algorithm
achieves the best result for the two datasets.  In particular, it is
substantially better than uniform sampling for solar radiation
data. For temperature data, since $\Theta(\mPhi^t\mPsi^t)$ of the
uniform sampling strategy is already near the lower bound\footnote{The
  lower bound of $\Theta(\mPhi^t\mPsi^t)$ is $\gamma=M/N$ if and only
  if $\mPhi^t\mPsi^t$ is a basis.}, the greedy algorithm provides
little improvement.  In the next section, we demonstrate how these
improvements translates into better reconstruction performance for
DASS.

\begin{table}[!t]
\renewcommand{\arraystretch}{1.2}
\caption{Average $\Theta(\mPhi^t\mPsi^t)$ achieved by different sampling scheduling methods ($\gamma=10\%$, SNR of the measurement=30dB)}
\label{condtab}
\centering
{\footnotesize\begin{tabular}{m{1.1in}|m{0.38in}|m{0.4in}|m{0.35in}}
\hline
\backslashbox[1.in]{\dataA{}}{Method} & uniform & random & Alg.~\ref{alg:greedy}\\
\hline
Temperature & 0.56 & 4.9$\times10^{15}$ & 0.54  \\
Solar radiation & 4.5$\times10^5$ & 1.8$\times10^{15}$ & 0.97 \\\hline
\end{tabular}}
\end{table}

\begin{figure}[!t]
\centering
\includegraphics[height=0.15\newhcol]{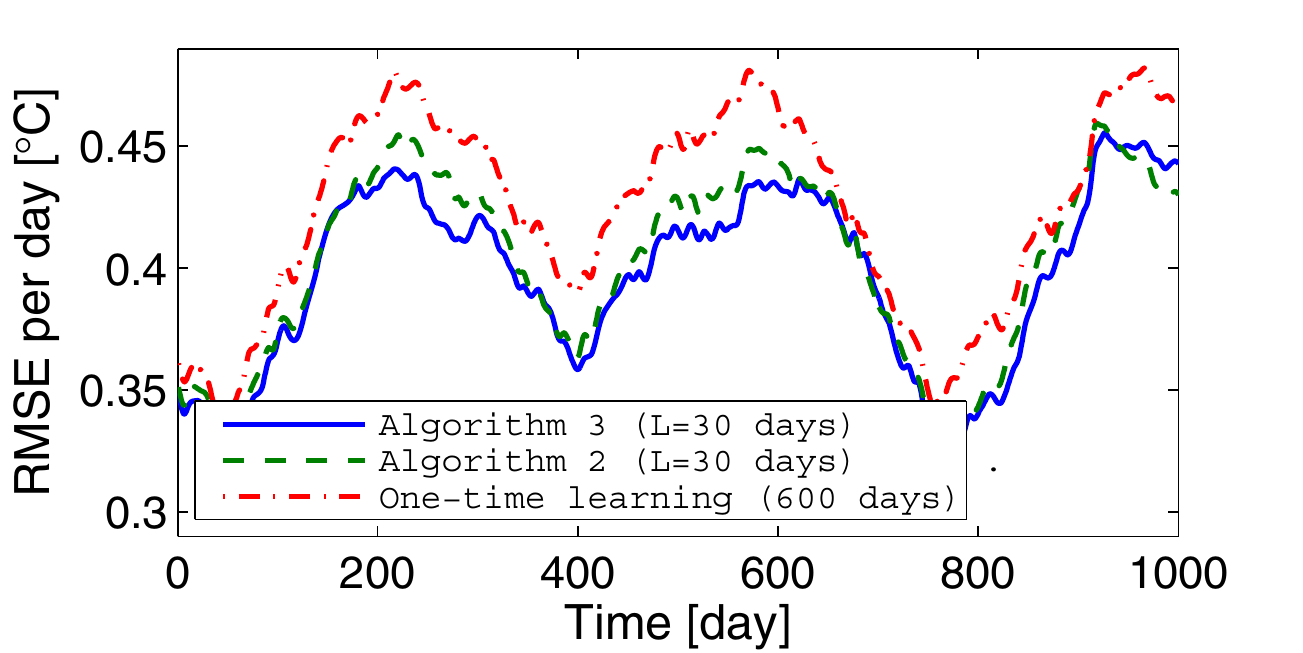}
\caption{Learning curves of DASS (\dataA{}-temperature, $\gamma=10\%$,
  SNR of the measurement=30dB): Comparisons of two online learning algorithms and a
  one-time learning algorithm with long backlog of past data. Note
  that Algorithm~\ref{updater2} achieves always the lowest error. }
\label{plot3}
\end{figure}

\keyitem{Learning Over Time} DASS is designed to learn the signal
statistics from past data. In practical scenarios, a long backlog of
data is usually infeasible and thus DASS should be designed to learn
the model from scratch. We proposed Algorithm~\ref{updater1} and
Algorithm~\ref{updater2} for this task. Figure~\ref{plot3} shows the
learning curves of these two algorithms over three years of data. As a
benchmark, we considered an offline method that learns the model from
600 days of past data and is represented by the red-dotted curve.

Note how Algorithm~\ref{updater1} and Algorithm~\ref{updater2} capture
the signal statistics precisely. In particular, it is interesting to
note that even if they use less data---the last 30 days---they are
generally better than the offline method that considers 600 days of
data. This phenomenon is due to the non-stationarity of the signal
model $\mPsi^t$ that is captured only by adaptive on-line
algorithms. Moreover, it is also clear that Algorithm~\ref{updater2}
with incremental PCA performs better than the buffer-based
Algorithm~\ref{updater1}.

In the following experiments, we will only consider
Algorithm~\ref{updater2} due to its better performance and lower
memory requirements.

\begin{figure}[!t]
\centering
\subfloat[]{\label{plot41}\includegraphics[height=0.21\newhcol]{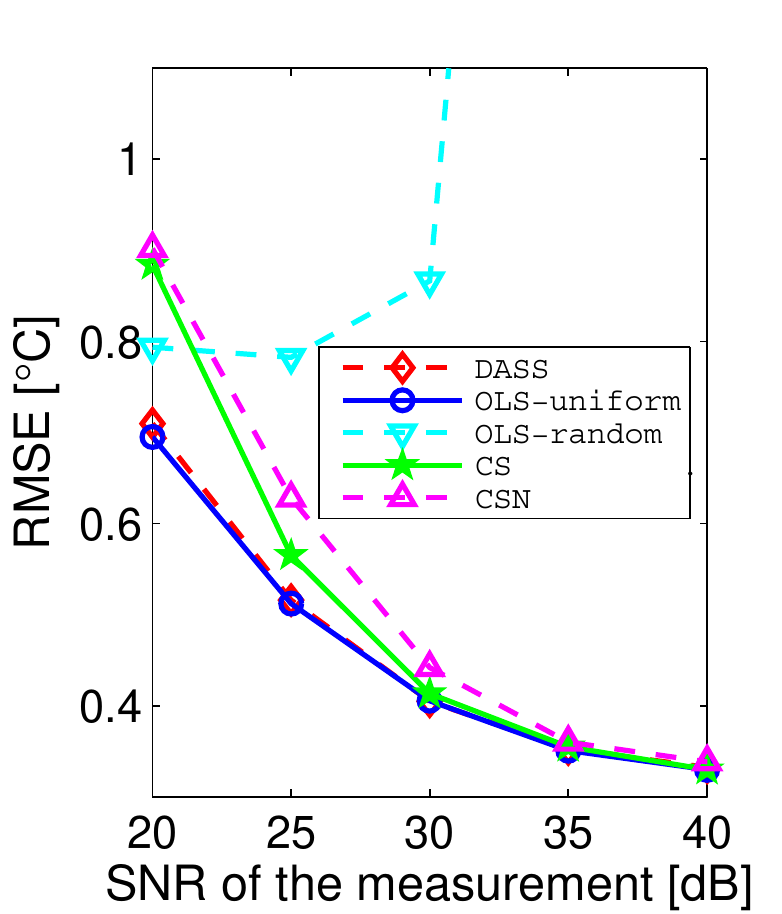}}
\subfloat[]{\label{plot42}\includegraphics[height=0.21\newhcol]{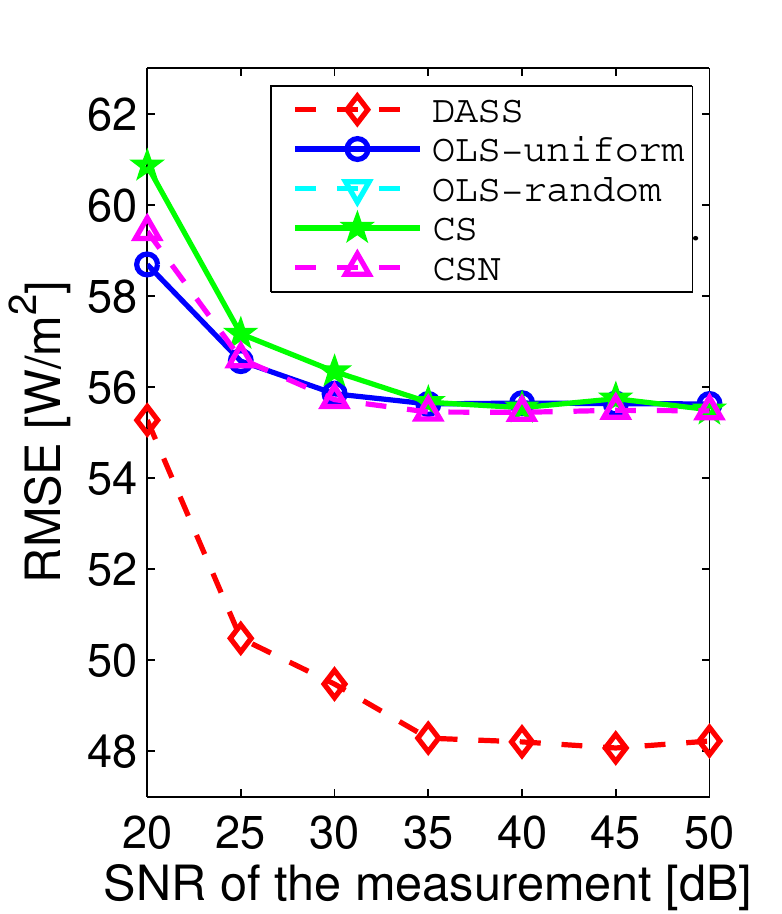}}
\caption{Reconstruction error (RMSE) w.r.t. SNR of the measurement, of DASS, OLS-uniform,
  OLS-random, CS and CSN, respectively ($\gamma=10\%$). The SNR is assumed to be accurately estimated. (a)
  \dataA{}-temperature. (b) \dataA{}-solar radiation. DASS is either
  on par with the best method, see (a), or significantly better, see
  (b). Note that in (b) OLS-random is not visible in the plot because
  it is significantly worse than the other methods. }
\label{plot4}
\end{figure}

\subsection{DASS versus Baseline Methods}
\label{sec5.2}
Here, we compare DASS with the baseline methods introduced in
Table~\ref{baselines}, namely, CS, CSN, OLS-random, and OLS-uniform.

\keyitem{Known Noise Level} For DASS, we need to choose the optimal
$K$ according to the cross-validation studied in
Figure~\ref{plot1}. Hence, we need to know the SNR of the
measurement. A similar parameter tuning is necessary for CSN, where
$\xi$ in Problem \eqref{eq4.2} represents the noise level.  Therefore,
whenever we consider the case of noisy measurements, an estimate of
the SNR of the measurement is necessary to avoid degradations of the reconstruction
quality.

In the first experiment, we assume that the estimation of the SNR is
exact. Figure~\ref{plot4} shows the comparison results of DASS,
OLS-uniform, OLS-random, CS and CSN, for both temperature and solar
radiation data. First, note that OLS-uniform generally performs better
than the two CS-based schemes, especially in low SNR regime. In high
SNR regime ($>35$dB), OLS-uniform, CS and CSN tend to perform the
same.  Second, the bad performance of OLS-random indicates that random
sampling is not a valid sampling strategy for neither temperature nor
solar radiation signals. Third, while DASS and OLS-uniform performs
almost equivalently for temperature data, we can note that DASS is
substantially better for solar radiation data. This fact is in
accordance with the analysis of $\Theta(\mPhi^t\mPsi^t)$ given in
Table~\ref{condtab}: if $\Theta(\mPhi^t\mPsi^t)$ due to uniform
sampling is large, then the sampling scheduling algorithm of DASS
(Algorithm~\ref{alg:greedy}) significantly improves the effectiveness
of sensing while preserving the average sampling rate.

\keyitem{Error in Noise Estimation} In practice, the estimation of the
noise level might be not exact. Here, we study the performance deviation of the
considered algorithms when there is an error in such estimates. More
precisely, we fix all the parameters and we vary the estimation error
of the SNR and then measure the performance of the algorithms in terms
of RMSE.

\begin{figure}[!t]
\centering
\includegraphics[height=0.15\newhcol]{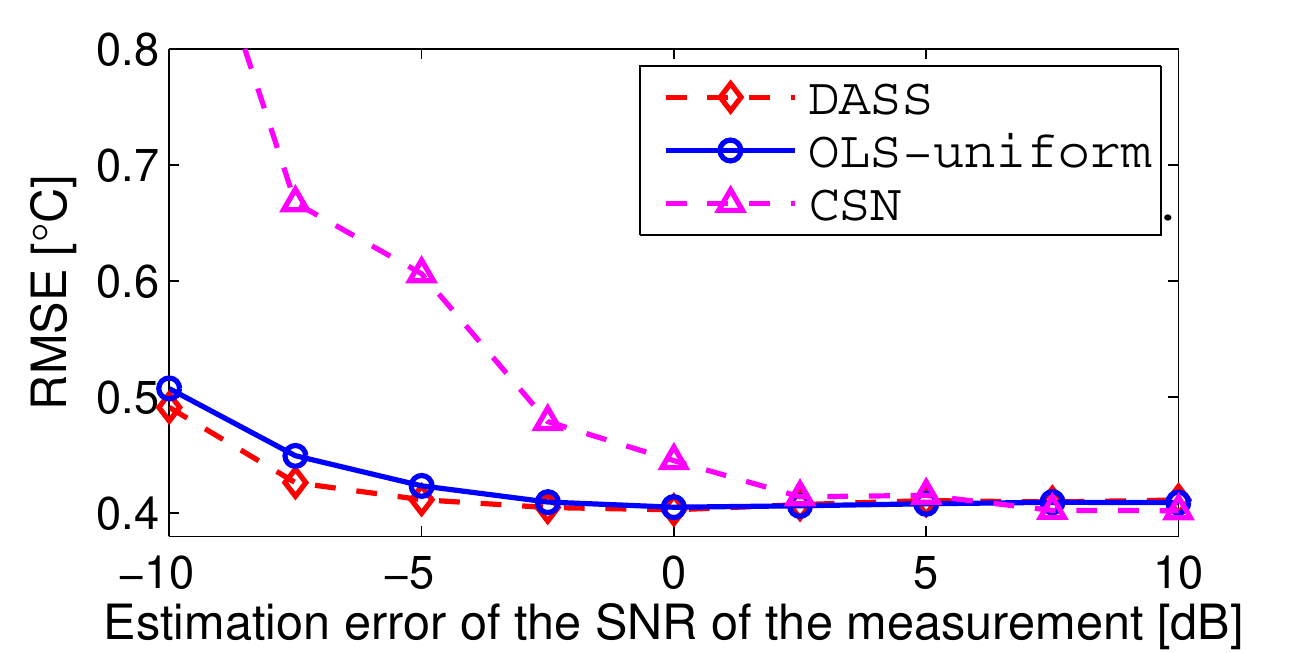}
\caption{Reconstruction error (RMSE) w.r.t. estimation error of the SNR of the measurement,
  of OLS-uniform, DASS and CSN, respectively (\dataA{}-temperature,
  $\gamma=10\%$). The true SNR is 30dB. Note that the proposed method
  is more robust to errors in the estimation of the noise
  power, when compared to other methods. }
\label{plot5}
\end{figure}

Figure~\ref{plot5} shows the reconstruction error with respect to the
estimation error of SNR, whereas the true SNR is 30dB. We can see that
DASS performs the best, and generally DASS and OLS-uniform are both stable w.r.t. errors in
the SNR estimation. However, the performance of CSN degrades severely
when the SNR is underestimated.

According to results given in Figure~\ref{plot4} and
Figure~\ref{plot5}, DASS is both more \emph{accurate} and
\emph{robust} when compared to the state-of-the-art sparse sensing
methods.

\subsection{DASS on Multiple Sensor Nodes}
\label{sec5.3}

As discussed in Section~\ref{sec2}, the concept of DASS can be extended
to multiple sensor nodes by concatenating the collected samples in a
single vector $\vy$ and using the same strategy as for the
single-node case.

Merging the data of all the spatial nodes possibly augments the
correlation; DASS may exploits such correlation to reduce the sampling
rate. In fact, if all the measurements collected by the sensors are
linearly independent then DASS generates the same sampling scheduling
that would have been optimized for each sensor individually. However,
if there exists some correlation between the different sensor
nodes, then DASS jointly optimizes the sensor scheduling so that the
total average sampling rate is reduced.

We denote by \emph{Joint DASS} the scheme that jointly reconstructs
the signals of the WSN (Figure~\ref{app3}), and \emph{Independent
  DASS} the scheme that independently reconstructs the signals of each
node.  Note that in both schemes, sensor nodes are operating in a
purely distributed manner; the difference is that \emph{Joint DASS}
aggregates the sensed data of all nodes and jointly processes them.

\begin{figure}[!t]
\centering
\includegraphics[height=0.16\newhcol]{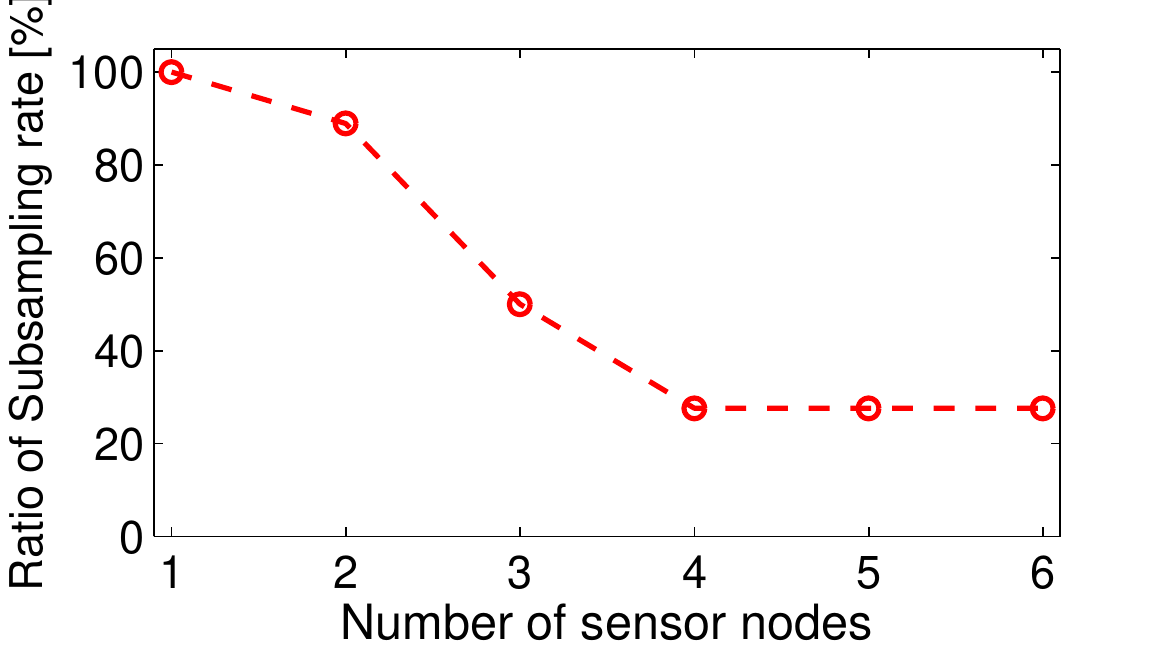}
\caption{Ratio of sampling rate between \emph{Joint DASS} and
  \emph{Independent DASS}, such that both schemes have the same
  reconstruction error (\dataB{}, SNR of the measurement=20dB). Note that the joint
  scheme always reduces the number of samples required, this is due to
the spatial correlation available in the sampled data.}
\label{plot6}
\end{figure}

Figure~\ref{plot6} shows the ratio between the subsampling rates of
\emph{Joint DASS} and \emph{Independent DASS}, using the data-set \dataB{}.  We can
see that as the number of sensor nodes increases, the required
sampling rate of \emph{Joint DASS} also gradually decreases. In
particular, with 4 nodes we can reduce the number of samples by 70\%
with \emph{Joint DASS}. Therefore, exploiting the spatial correlation
further enhances the energy reduction of DASS.  On the other hand, the
benefit flatten out when we consider 5 or more sensor nodes. The
intuition behind this phenomenon is that the last two sensor
nodes are far apart from the others and there is no more correlation to
exploit, see the rightmost two nodes in Figure~\ref{map}.

\begin{figure}[!t]
\centering
\includegraphics[height=0.16\newhcol]{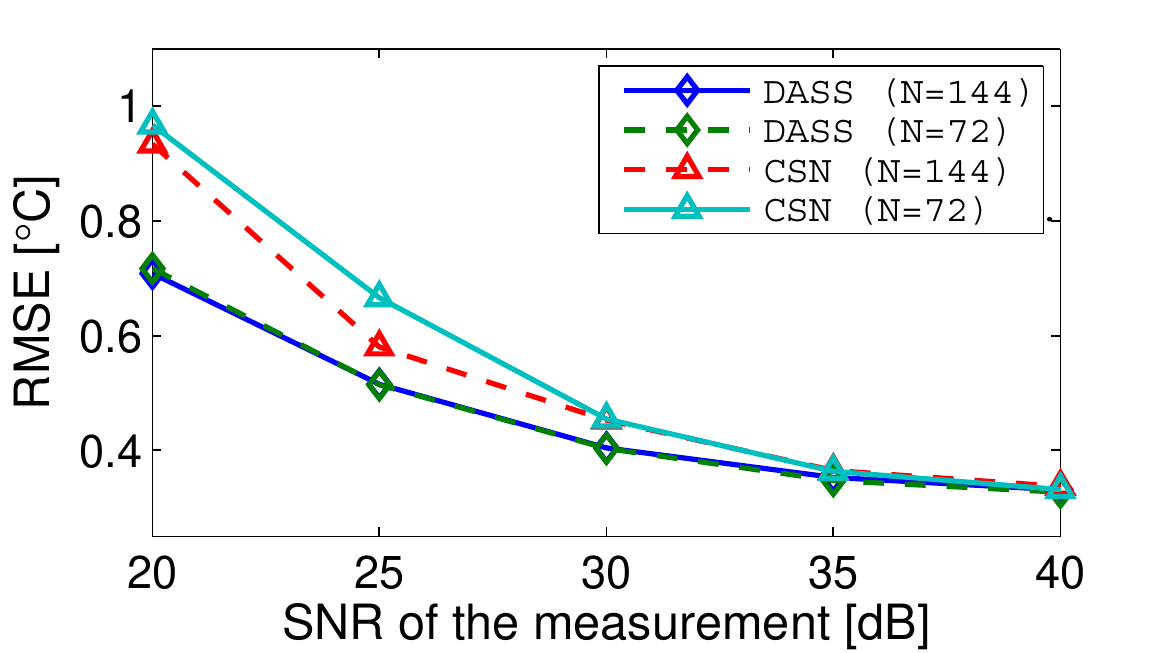}
\caption{Reconstruction error (RMSE) of DASS and CSN, when block
  length $N=72$ or 144 (\dataA{}-temperature, $\gamma=10\%$). Note
  that one day has 144 data points so $N=72$ is half of the day.
  The performance of DASS is only slightly affected by a change of $N$,
  while CSN is considerably affected in the low SNR regime. }
\label{plot7}
\end{figure}

\subsection{Blocks with Weaker Correlation}
\label{sec5.4}
In real applications, the block length $N$ must be chosen such that
the delay of the WSN respects the design specification while the
correlation between blocks is maximized. In all experiments above,
$N$ is chosen so that one block represents one day, which intuitively
fits signals with strong diurnal cycles, such as temperature signals.
In practice, it is essential to evaluate how DASS performs with a
sub-optimal $N$.  In this section, we use the same dataset
\dataA{}-temperature, but splitting one day into two blocks. This
means that we transmit and reconstruct signals two times per day and hence
the correlation between the different temporal blocks is
smaller. Figure~\ref{plot7} compares DASS and CSN with two possible
block length: a full day---$N=144$--- and half a day---$N=72$.  We can
note that the performance of DASS is only slightly affected by the smaller block length,
while CSN is considerably affected in the low SNR regime.

\section{Energy Saving over Traditional Data Collection Schemes}
\label{sec6}

In Section~\ref{sec5}, we have shown that DASS achieves better
performance w.r.t. the state-of-the-art \emph{sparse sensing
  schemes}. In this section, we study the \emph{overall energy saving}
of DASS w.r.t. the \emph{traditional data collection
  schemes}~\cite{Sadler2006, Zordan2012}. The energy saving is
particularly significant on platforms where the energy consumed for
sensing is more pronounced. This is intuitive since DASS can
substantially reduce the number of sensing samples. Nevertheless, our
analysis shows that this saving is also noticeable on platforms with
small sensing cost, e.g. a \emph{Tmote-sky}
node~\cite{Werner-Allen2006}.

\begin{figure}[!t]
\centering
\includegraphics[height=0.15\newhcol]{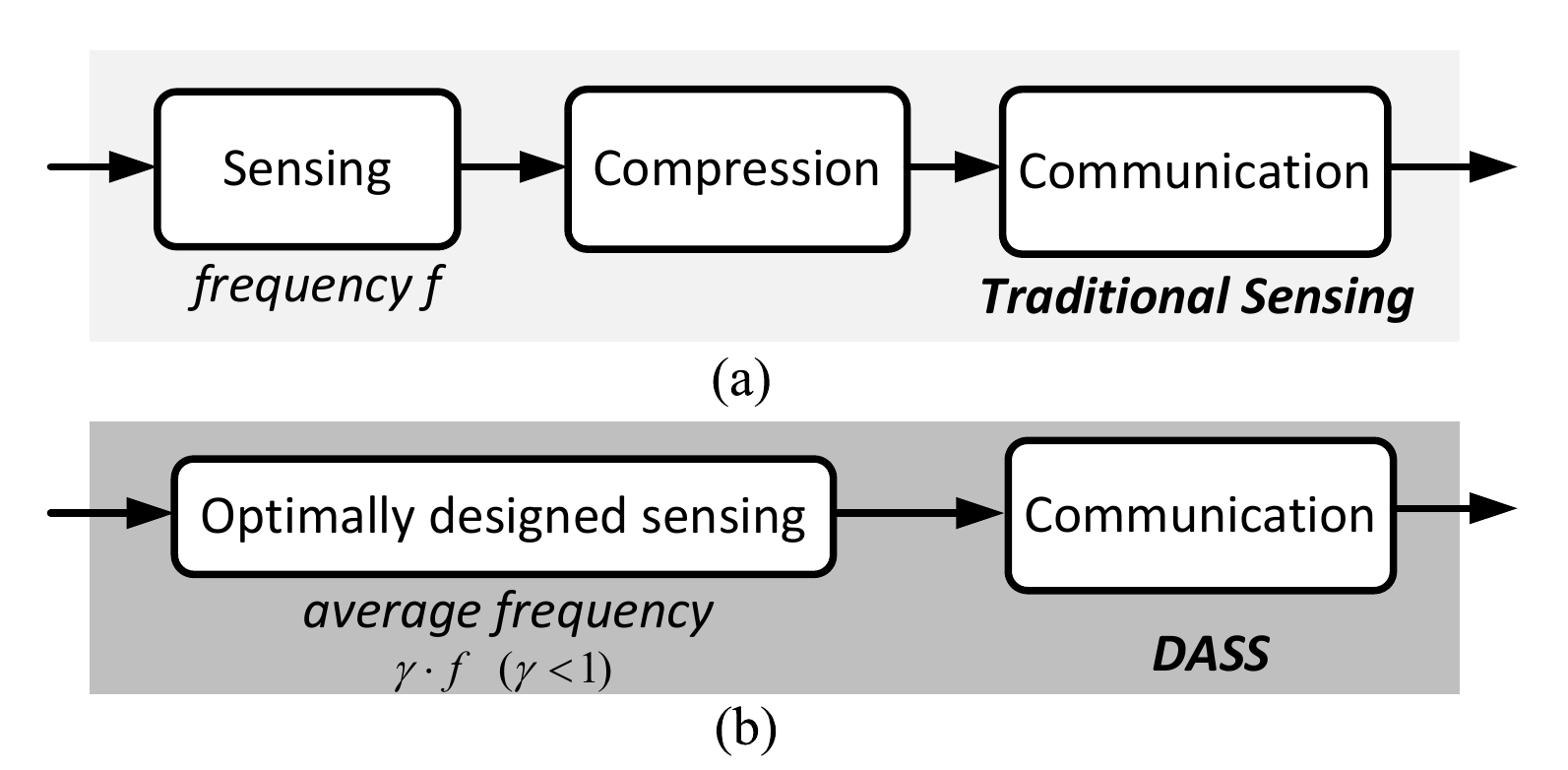}
\caption{Two approaches to sensing in a WSN node. (a) Traditional
  scheme: collect periodical samples at a frequency $f$, compress and
  transmit the compressed data. (b) DASS: collect samples with an
  optimized temporal pattern at an average frequency $\gamma\cdot f$
  and transmit the raw data.}
\label{sensing}
\end{figure}

The traditional data collection schemes typically sample the physical
field at a high frequency $f$ as in \eqref{eq:trad_sensing} and then
compress the samples to reduce the communication rate, see
Figure~\ref{sensing}a. In contrast, DASS collects measurements using
an optimized sampling pattern and a reduced average sensing frequency
$\gamma\cdot f$, where $\gamma<1$. Then, each sensor node transmits the
raw data points without any compression, see Figure~\ref{sensing}b. In
both traditional schemes and DASS, we aim at precisely reconstructing
the signal $\vx$.

\begin{figure}[!t]
\centering
\includegraphics[height=0.25\newhcol]{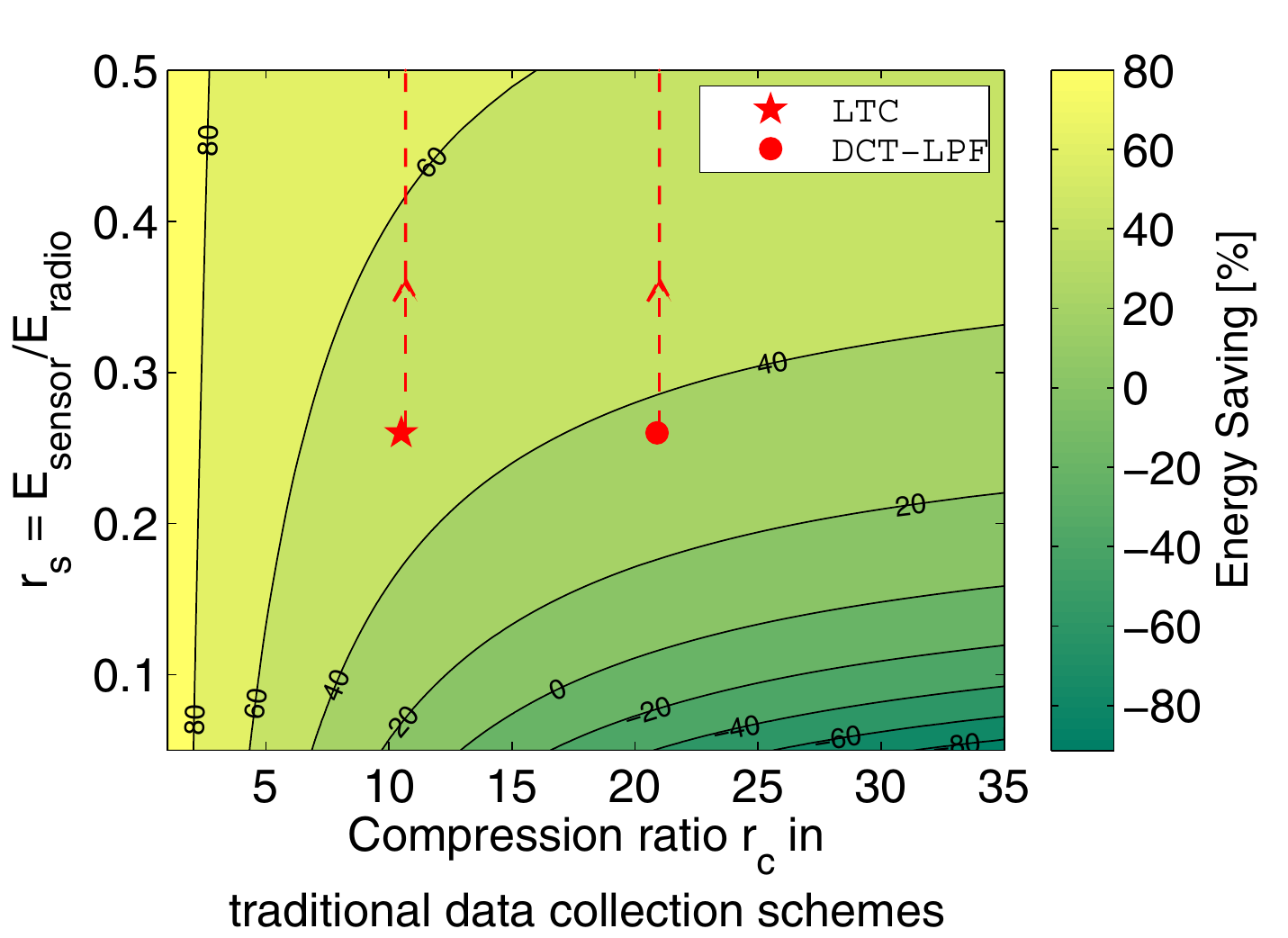}
\caption{Relative energy saving of DASS ($\gamma=10\%$)
  w.r.t. traditional data collection schemes.  The saving depends on
  the sensing platform (value of $\mathbf{r}_s$) and the compression
  ratio $\mathbf{r}_c$ in traditional sensing.  The ``star'' and ``circle'' markers
  represent the energy saving on \emph{Tmote-sky}, when DASS achieves
  the same reconstruction error as traditional sensing using LTC and
  DCT-LPF compression methods~\cite{Zordan2012} (on dataset
  \dataA{}-temperature) .  The dashed lines indicate further savings
  when $\mathbf{r}$ increases, that is for sensors with higher energy
  costs.}
\label{saving}
\end{figure}

\begin{figure}[!t]
\centering
\subfloat[]{\label{consumption1}\includegraphics[height=0.17\newhcol]{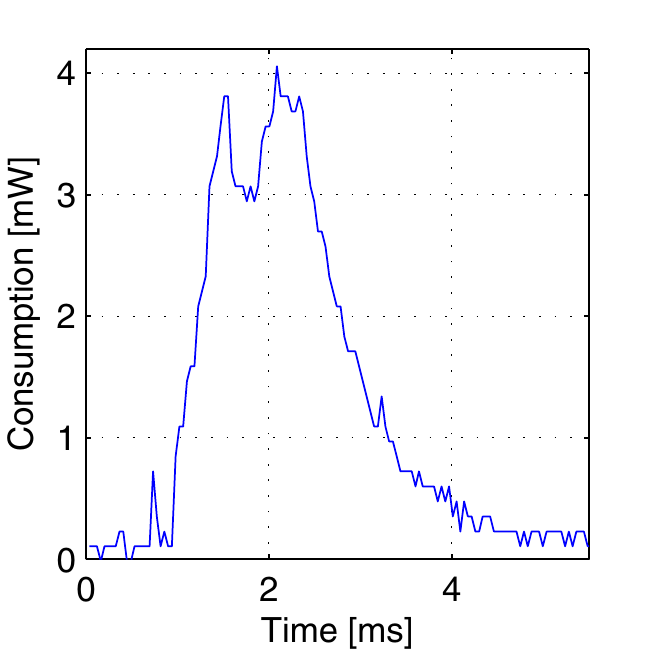}}
\subfloat[]{\label{consumption2}\includegraphics[height=0.17\newhcol]{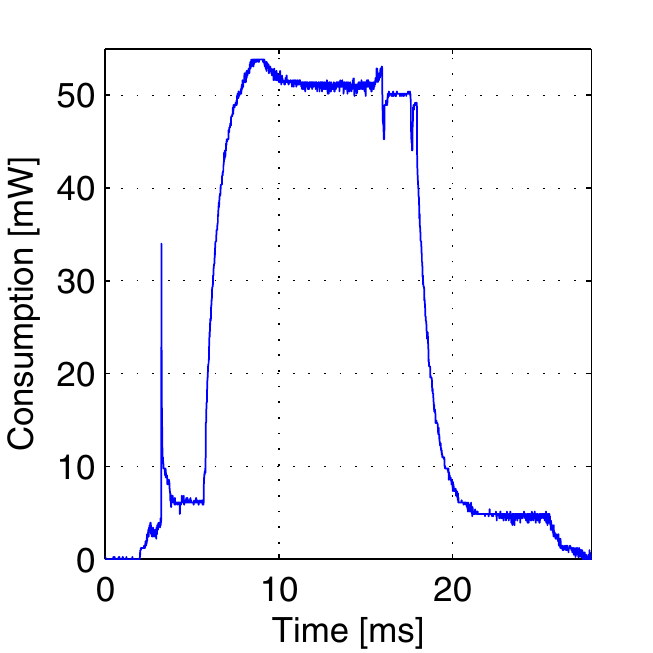}}
\caption{Energy consumptions of a \emph{Tmote-sky} sensor: (a) while
  the node measures one sample of light intensity (two-bytes),
  $E_{sensor}=7.5\times10^{-6}\textrm{J}$; (b) while the node
  transmits a packet with 24 bytes of payload, $24\cdot
  E_{radio}=6.9\times10^{-4}\textrm{J}$.}
\label{measuretmote}
\end{figure}

It is clear that DASS reduces the energy consumption for the sensing
operations over the traditional scheme.  However, DASS may not
necessarily consume less communication energy, since the compression
ratio $\mathbf{r}_c$\footnote{$\mathbf{r}_c$ equals uncompressed size /
  compressed size.} used in traditional sensing is generally better
than $1/\gamma$. In fact, existing data compression schemes can
achieve a compression ratio $\mathbf{r}_c$ of $1.5\sim 5$ for lossless
coding~\cite{Sadler2006}, and $5\sim 50$ for lossy
coding~\cite{Zordan2012}, while a typical value of $\gamma$ used in
DASS is $0.1$. Hence, there is a tradeoff between the energy saved on
sensing and communications.

Such tradeoff between the different energy consumption depends on
platform-specific parameters.  In particular, we denote the energy
consumption for collecting and transmitting one sample as $E_{sensor}$
and $E_{radio}$, respectively. An interesting figure is the ratio
between the two energy values, that we denote as
$\mathbf{r}_s=E_{sensor}/E_{radio}$.  Intuitively, the larger
$\mathbf{r}_s$, the larger the energy savings obtained by DASS. For
the traditional data collection schemes, we assume that the
compression step has a negligible energy cost. For DASS we use a
subsampling rate of $\gamma=0.1$, which means that 10\% of the
original signal is sampled and transmitted.

Under these assumptions, we can quantitatively analyze the relative energy
savings of DASS w.r.t. the traditional sensing as a 2-D function of
the platform parameter $\mathbf{r}_s$ and the compression ratio
$\mathbf{r}_c$ achieved by the compression stage of the traditional
scheme. Such function representing the energy saving is plotted in
Figure~\ref{saving}. We see that there is a line, indicated by the
zero value, that defines where DASS is more energy-efficient than the
traditional schemes. Above the line, a WSN consumes less energy if it
uses DASS and vice versa. Note that DASS is only less efficient in the scenarios where the compression ratio $\mathbf{r}_c$ is very high and the platform parameter $\mathbf{r}_s$ is very low.

We also looked at the energy savings for a plausible real world
scenario. More precisely, we consider \emph{Tmote-sky}, a low-power
sensing platform widely used in WSNs~\cite{Werner-Allen2006}; it has a
photodiode sensor that measures the light intensity of the
surroundings and can communication with others through short-range
radio.  We measured the two energy consumptions $E_{sensor}$ and
$E_{radio}$ of \emph{Tmote-sky} in a set of experiments, and an
example of the results is given in Figure~\ref{measuretmote}. In
particular, the experiments indicate that $\mathbf{r}_s=0.26$.  To
evaluate the energy consumption of a traditional scheme, we need to
choose a specific compression algorithm and measure the achieved
$\mathbf{r}_c$. Zordan et al.~\cite{Zordan2012} have recently compared
various lossy compression algorithms and showed that
DCT-LPF~\cite{Zordan2012} achieves the best performance in terms of
compression ratio. However, it is also a complex algorithm and may
have a significant energy consumption on a resource-limited platform
such as \emph{Tmote-sky}. Therefore, we also consider a lightweight
algorithm, LTC~\cite{Schoellhammer2004}, that achieves the lowest
energy consumption on WSN nodes if the energy cost for compression is
considered.

Here, we ignore the energy cost of compression and we compare both
algorithms with DASS. Note that, if we consider computational energy
cost, the benefit of DASS will be even larger since it requires
minimal on-board computation.  We implement and evaluate the two
algorithms on the dataset \dataA{}-temperature, and record the
corresponding compression ratio $\mathbf{r}_c$ when their
reconstruction errors are the same as those achieved by DASS.

The ``star'' and ``circle'' markers in Figure~\ref{saving} show the
energy savings of DASS over a \emph{Tmote-sky} that compresses the
data with LTC and DCT-LPF, respectively. The energy savings for the
two cases are equal to 50\% and 35\% and go up to 60\% if
$\mathbf{r}_s$ increases due to a higher energy cost for sensing, as
denoted by the dashed lines in Figure~\ref{saving}. This scenario
could be realistic for many WSNs, in particular those using sensor
belonging to the following two classes:
\begin{itemize}
\item Sensors with high energy consumption: for example an air
  pollution sensors consume $30\sim 50$ mW instead of the 3 mW of a
  \emph{Tmote-sky}'s light sensor.
\item Sensors with long sampling time: for example the anemometer, a
  sensor that measures wind's direction and strength, requires $1\sim
  3$ seconds of continuous measurement per sample instead of the 4 ms
  of the \emph{Tmote-sky}'s light sensor.
\end{itemize}

\section{Conclusions}
\label{sec8}

In this paper, we proposed DASS, a novel approach for sparse sampling
that optimizes sparse sampling patterns for precisely recovering
spatio-temporal physical fields.  DASS is based on three main
blocks. First, it adaptively learns the signal statistics from past
data. Second, it dynamically adjusts the sampling pattern according to
the time--varying signal statistics. Third, it recovers the signal
from the limited amount of collected samples and according to the
learnt signal statistics.

We demonstrated the effectiveness of DASS through extensive
experiments using two real-world meteorological datasets. The results
show significant improvements over the state-of-the-art methods. These
improvements are more pronounced in the presence of significant
spatial and/or temporal correlation in the sampled data by WSN.

We evaluated DASS on static WSNs; however, DASS is flexible
and can be applied to other sensing scenarios such as mobile WSNs. For
instance, sensors are installed on top of buses for collecting
various environmental data along their
trajectories~\cite{Aberer2010}. The collected samples show strong
correlation due to the fixed route periodically taken by the
buses. In future work, we will analyze the advantages of an optimized sensing
schedule in such cases, where the constraint is not the energy
consumption but the relatively slow speed of sampling of certain
pollution sensors.

{
\bibliographystyle{abbrv}
\bibliography{IEEEabrv,biblio2,papers}  
}

%
%
\end{document}